\documentclass
[twocolumn,aps,prc,amsmath,amssymb,floatfix]
{revtex4-1}
\usepackage{CJK}%,pinyin}            % chinese fonts
\usepackage{graphicx}         %
\usepackage{bm}                      % bold math
\usepackage{mathptmx}                % math+times fonts
\usepackage{dcolumn}                 % Align table columns on decimal point
\usepackage{multirow}                % tables
\usepackage{xcolor}                  %
\usepackage{amssymb}
\usepackage{amsfonts}
\usepackage{slashed}
\usepackage{mathrsfs}
\usepackage{soul}
\usepackage{cancel}
\usepackage[
breaklinks,pdfstartview=FitH,CJKbookmarks=true,
bookmarksnumbered=true,bookmarksopen=true,pdfborder={0 0 1},
colorlinks=true,linkcolor=blue,urlcolor=blue,anchorcolor=blue,citecolor=blue
            ]{hyperref}
\begin{document}

\def\be{\begin{equation}} \def\ee{\end{equation}}
\def\bal#1\eal{\begin{align}#1\end{align}}
\def\bse#1\ese{\begin{subequations}#1\end{subequations}}
\def\rra{\right\rangle} \def\lla{\left\langle}
\def\rv{\bm{r}} \def\tv{\bm{\tau}} \def\sv{\bm{\sigma}}
\def\tt{(\tv_1\cdot\tv_2)} \def\ss{(\sv_1\cdot\sv_2)}
\def\non{\nonumber}
\def\ra{\rightarrow}
\def\arsinh{\mathop{\text{arsinh}}}
\def\al{\alpha}
\def\la{\Lambda}
\def\eps{\epsilon}
\def\ms{\,M_\odot}
\def\mmax{M_\text{max}}
\def\rmin{R_\text{min}}
\def\l14{\Lambda_{1.4}}
\def\r14{R_{1.4}}
\def\rd14{R^{(D)}_{1.4}}
\def\rn14{R^{(N)}_{1.4}}
\def\lr14{\l14 / \la_\text{fit}(\r14)}
\def\fm3{\;\text{fm}^{-3}}
\def\km{\;\text{km}}
\def\mev{\;\text{MeV}}
\def\gev{\;\text{GeV}}
\def\mfm{\text{MeVfm}^{-3}}
\def\fb#1{\textcolor{blue}{#1}}
\def\hcd#1{\textcolor{magenta}{#1}}
\long\def\hj#1{\color{red}#1\color{black}}
\def\OFF#1{}
\long\def\add#1{\color{blue}#1\color{black}}
\newcommand{\del}[1]{{\color{red}\st{#1}}\color{black}}
\title{
Is XTE J1814-338 a dark matter-admixed neutron star?
}

%\begin{CJK*}{UTF8}{gbsn}

\author{Hong-Ming Liu$^{1}$}\email[]{liuhongming13@126.com}
\author{Peng-Cheng Chu$^{1}$}\email[]{kyois@126.com}
\author{\hbox{He Liu$^{1}$}} \email[]{liuhe@qut.edu.cn}
\author{Zeng-Hua Li$^2$} \email[]{zhli09@fudan.edu.cn}
\author{Xiao-Hua Li$^{3,4}$} \email[]{lixiaohuaphysics@126.com}

\affiliation{
$^1$\hbox{
The Research Center for Theoretical Physics, Science School, Qingdao University of Technology,}\\
{Qingdao 266033, People's Republic of China}\\
\hbox{$^2$ Institute of Modern Physics, Key Laboratory of Nuclear Physics and Ion-Beam Application, MOE, Fudan University,} \\ 
{Shanghai 200433, People's Republic of China}\\
\hbox{$^3$ School of Nuclear Science and Technology, University of South China, Hengyang,
421001, People's Republic of China}\\
\hbox{$^4$ Cooperative Innovation Center for Nuclear Fuel Cycle Technology \& Equipment,University of South China,}\\
{Hengyang, 421001, People's Republic of China}
}

\date{\today}

\begin{abstract}
We explore the possibility that the ultracompact object XTE J1814-338 is a dark matter-admixed neutron star (DNS) 
based on a general-relativistic two-fluid approach, in which one fluid describes ordinary nuclear matter and the other dark matter. 
The equations of state of nuclear matter and dark matter are described by the Brueckner-Hartree-Fock (BHF) 
theory with realistic three-body forces and a generic bosonic self-interacting dark matter model, respectively.
Based on the observational data, we determine the parameter space for XTE J1814-338 as a DNS, 
and find that only a halo stellar configuration with a high dark matter fraction is allowed.
Furthermore, we predict other observable properties of XTE J1814-338 and provide robust lower bounds for these quantities.
\end{abstract}

\maketitle
%\end{CJK*}

%-------------------------------------------------------------------------------
\section{Introduction}

Neutron stars (NSs), as the densest objects known in the Universe, 
provide an unique environment to study cold and dense matter under extreme conditions. 
With central densities exceeding several times the nuclear saturation density, 
NSs serve as natural laboratories to explore the equation of state (EOS) 
of asymmetric nuclear matter at supranuclear densities \cite{Steiner05,Burgio21}. 
Recent advances in multi-messenger astronomy have brought significant progress in constraining the EOS. 
The detection of gravitational wave (GW) signal from binary neutron
star (BNS) merger GW170817 has imposed stringent bound on tidal deformability \cite{Abbott17,Abbott18}, 
while precise mass measurements of heavy pulsars such as PSR J0740+6620 with mass $M$ = 2.08 $\pm$ 0.07$\ms$ 
set stringent requirements on the maximum mass supported by the EOS \cite{Cromartie20}. 
In addition, X-ray pulse-profile modeling with Neutron Star Interior Composition Explorer (NICER)
space mission has provided simultaneous mass and radius measurements
for PSR J0030+0451 \cite{Miller19,Riley19}, PSR J0740+6620 \cite{Riley21,Miller21} and PSR J0437-4715 \cite{Choudhury24}. 
EOSs that simultaneously satisfy these astronomical constraints are considered reliable. 
However, several peculiar astrophysical observations such as 
extraordinary compact or massive objects remain difficult to reconcile within the framework of conventional NS models 
\cite{Romani22,Doroshenko22,Abac24,Ewan24}. 
These tensions suggest the emergence of additional degrees of freedom 
or new physics beyond conventional baryonic matter inside NSs. 
Among various possibilities, 
the admixture of dark matter (DM) provides a compelling scenario that could naturally account for such anomalies 
\cite{Routaray23,Ren24,Barbat24,Liu26}.

Currently, 
the possibility that DM may play a role in NS structure has attracted growing interest. 
Robust astrophysical and cosmological evidence 
from galactic rotation curves to cosmic microwave background anisotropies 
indicates that DM contributes roughly 27\% of the total energy density of the Universe \cite{Bertone05,Massey10}. 
However, 
its particle nature and interactions with ordinary matter remain elusive \cite{Begeman91,Wittman00, Abdalla10}. 
To characterize its properties, 
a range of experimental approaches have been employed. 
Direct detection experiments, 
such as XENONnT \cite{Aprile22}, DarkSide-50 \cite{Agnes18}, and PandaX-II \cite{Wang20}, 
aim to measure nuclear recoils induced by DM scattering. 
Indirect detection experiments, 
including AMS-02 \cite{Accardo14}, Fermi-LAT \cite{Atwood09} and CALET \cite{Adriani18}, 
search for high-energy gamma rays, cosmic rays, or neutrinos that may originate from DM annihilation or decay. 
Meanwhile, collider searches, most notably at the Large Hadron Collider (LHC) \cite{Sato16}, 
attempt to produce DM candidates in high-energy collisions, 
identified through missing transverse momentum signatures. 
These complementary efforts have significantly narrowed the viable parameter space for DM models, 
yet no conclusive evidence of DM particles has been obtained.

In this context, 
NSs emerge as a natural astrophysical laboratory for probing DM properties in a way complementary to terrestrial searches 
\cite{Shakeri24,Barbat24,Kumar24}. 
If DM interacts weakly with baryons, 
NSs can efficiently capture and accumulate DM during their formation and subsequent evolution \cite{Bell21,Collier22,Gleason22}. 
Even a modest DM component may alter their macroscopic properties, 
including the mass-radius relation, tidal deformability, 
stability profiles and cooling rates giving rise to the so-called DM-admixed compact stars 
\cite{Xiang14,Tolos15,ADas19,Yang21,Kain21,Emma22,Bauswein23,Routaray23b,Liu23,Wu25,Shakeri24,Liu24,Sedrakian19}. 
Importantly, 
these potential signatures maybe observable in the multi-messenger observations, 
making NSs a promising indirect detection window into the fundamental nature of DM \cite{Zhang22,Gupta23}.

Recently, 
the ultracompact object XTE J1814-338, 
with an inferred mass and radius of $M$ = 1.21 $\pm$ 0.05$\ms$ and $R$ = 7.0 $\pm$ 0.4$\km$ at 68\% confidence level (CL), 
has attracted considerable attention \cite{Kini24}. 
Its reported mass-radius combination is challenging to reconcile with predictions from conventional hadronic EOSs, 
implying that additional degrees of freedom or exotic components may be required to explain its nature.
Ref.~\cite{Pitz25} suggests that XTE J1814-338 could be a bosonic star containing a nuclear matter core.
Ref.~\cite{Yang25} proposes that its structure could correspond to a strange star admixed with mirror DM.
Moreover, Ref.~\cite{Lopes25} indicates that XTE J1814-338 may be interpreted as a fermionic DM-admixed NS (DNS), 
and further infers the corresponding physical properties of this configuration.

In this work, 
we systematically investigate the structure of XTE J1814-338 within the framework of DNSs. 
We aim to assess the viability of this interpretation by exploring how 
the presence of DM affects the global properties of the star across the full parameter space. 
Furthermore, 
we present theoretical predictions for two key observables including the tidal deformability and moment of inertia, 
which could serve as discriminating probes in future observations. 
Our results offer a potential resolution to the puzzle of XTE J1814-338 
and contribute to the broader effort to constrain both the EOS of dense matter and the properties of DM.

This article is organized as follows.
In the next section~\ref{Sec.2}, the EOSs of nuclear matter and DM used in this work are briefly described.
The detailed calculations and discussion are presented in Sec.~\ref{Sec.3}.
A summary is given in Sec.~\ref{Sec.4}. 
Natural units ($G=c=\hbar=1$) are used throughout this article.

%-------------------------------------------------------------------------------
\section{Formalism}
\label{Sec.2}

%-------------------------------------------------------------------------------
\subsection{Equation of state for nuclear matter}
%BHF model
The nuclear EOS employed in our calculations is derived within the Brueckner-Bethe-Goldstone (BBG) theory, 
{\add{in}} which the key quantity is the interaction matrix $K$, and satisfies the following equation:
\begin{equation}
K(\rho, x; E) = V + \mathrm{Re} \sum_{1,2} 
V  \frac{|12\rangle Q \langle12|}{E - e_1 - e_2} K(\rho, x; E),
\label{eq:1}
\end{equation}
where $V$ is the bare nucleon-nucleon interaction, 
$E$ is the starting energy, and the mulyi-indices 1 and 2 denote momentum, spin, and isospin. 
In the present calculations, 
we adopt the Argonne $V_{18}$ (V18) potential \cite{Wiringa95} as a realistic $V$ 
supplemented by a consistent meson-exchange three-nucleon force, 
which enables a correct reproduction of the nuclear matter saturation point~\cite{Grange89,Zuo02,Li08a,Li08b}
 and other bulk properties around saturation density~\cite{Wei20,Burgio21b}.
The proton fraction is defined as $x_p = \rho_p / \rho$, 
with $\rho_p$ and $\rho$ the proton and total baryon densities, respectively.
and the single-particle (s.p.) energy reads
\begin{equation}
e_1 = \frac{k_1^2}{2m_1} + U_1,
\label{eq:2}
\end{equation}
where $k_1$ and $m_1$ are the momentum and mass of s.p. state, respectively.
The BHF approximation for the s.p. potential using the continuous choice is expressed as
\begin{equation}
U_1(\rho, x_p) = \mathrm{Re} \sum_{k_2 < k_F} 
\langle 12 | K(\rho, x_p; e_1 + e_2) | 12 \rangle_A,
\label{eq:3}
\end{equation}
where $k_F$ is the Fermi momentum, the subscript $A$ denotes antisymmetrization of the matrix elements. 
Due to the appearance of the s.p. potential $U(k)$ in Eq.~(\ref{eq:2}), 
the coupled Eqs.~(\ref{eq:1}) to (\ref{eq:3}) must be solved self-consistently. 
The corresponding BHF energy density is given by
\begin{equation}
\varepsilon_{nucl} = 
\sum_{i = n,p} 2 \sum_{k < k_F^{(i)}}
\left(
\frac{k^2}{2m_i}
+ \frac{1}{2} U_i(k)
\right),
\label{eq:energy_density}
\end{equation}
and all other thermodynamic quantities can be derived in a consistent manner. 
In particular, the chemical potentials and pressure are obtained as
\begin{equation}
\mu_i = \frac{\partial \varepsilon_{nucl}}{\partial \rho_i},
\label{eq:mu}
\end{equation}
\begin{equation}
p_{nucl} = \rho^2 \frac{\partial (\varepsilon_{nucl} / \rho)}{\partial \rho}
= \sum_i \mu_i \rho_i - \varepsilon_{nucl}.
\label{eq:pressure}
\end{equation}

In this paper, we focus exclusively on $\beta$-stable stellar matter
composed of nucleons and leptons (electrons $e$ and muons $\mu$). 
At a given baryon density $\rho$,
the composition is determined by imposing both $\beta$-equilibrium 
and local charge neutrality
\bal
&\mu_n - \mu_p = \mu_e = \mu_{\mu},\\
&\rho_p = \rho_e + \rho_{\mu},
\eal
where leptons are treated as non-interacting Fermi gases.
The total energy density $\eps_N$ and pressure $p_N$ are then 
obtained by adding the leptonic contributions to the
nucleonic ones.

We mention that the BHF approach provides only the EOS for the uniform matter region
($\rho \gtrsim 0.08\,{\rm fm}^{-3}$), while the low-density crust is
described by the Shen20 EOS \cite{Shen20}.
The mass-radius relation of a pure NS can then be obtained 
by solving the well-known Tolman-Oppenheimer-Volkoff (TOV) equations~\cite{Tolman39,Oppenheimer39}.

\begin{figure*}[t]%..............................................................
\centering
\includegraphics[width=\textwidth]{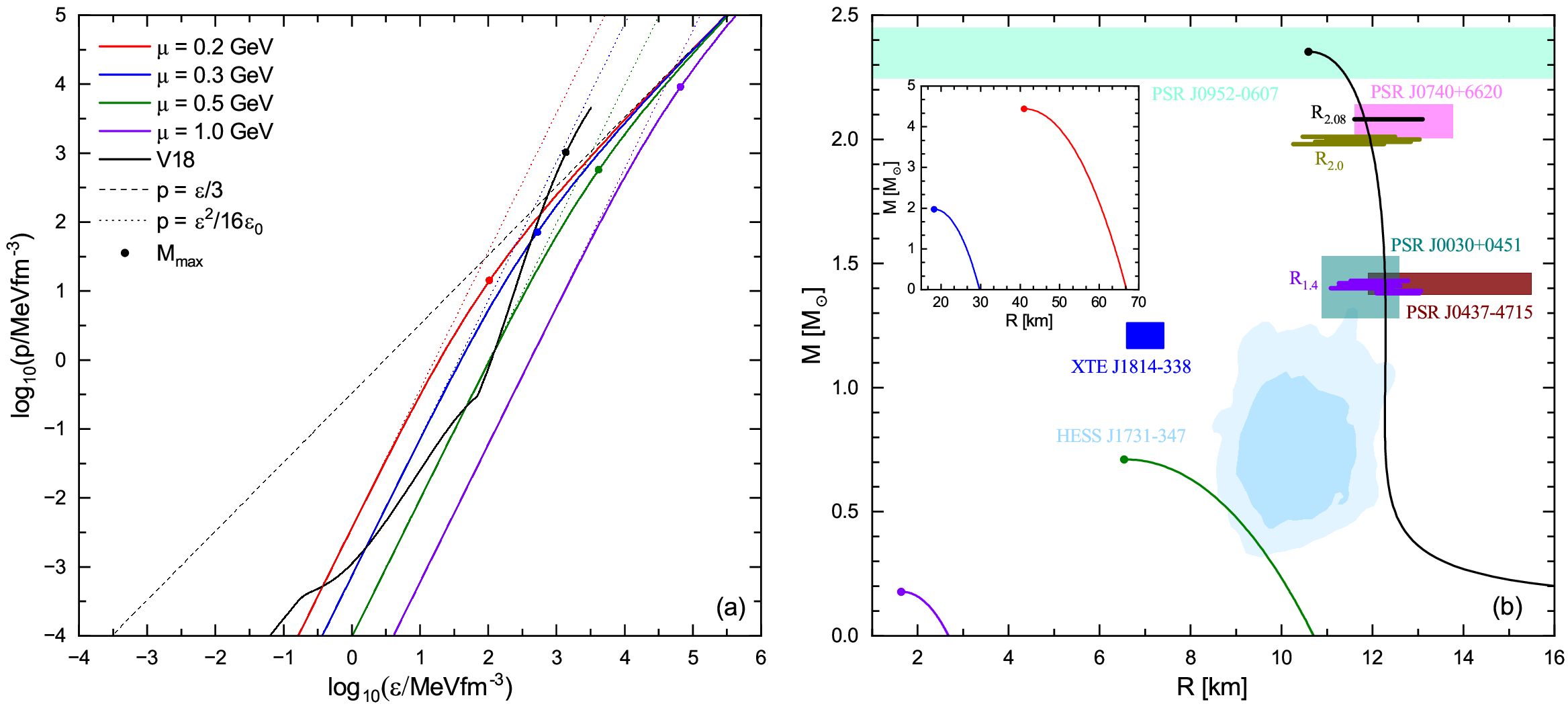}
\caption{
(a) EOSs of NS matter (V18 EOS) and pure self-interacting
bosonic DM stars matter with different DM particle masses {\add{$\mu$}}.
The circles indicate the values of the maximum-mass configurations.
The low-density and high-density asymptotics from Eqs.~(\ref{dmeq:2}) and (\ref{dmeq:3}) 
are indicated by the dashed and dotted curves, respectively.
(b)
Corresponding mass-radius relations obtained from the single-fluid TOV
equations~\cite{Tolman39,Oppenheimer39}.
The shaded areas represent the constraints from
PSR J0952-0607 \cite{Romani25}, PSR J0030 + 0451 \cite{Vinciguerra24}, 
PSR J0740 + 6620 \cite{Salmi24}, PSR J0437-4715 \cite{Miller26}, HESS J1731-347 \cite{Doroshenko22},
and XTE J1814-338 \cite{Kini24}, as labeled in the panel. 
The horizontal bars labeled $R_{1.4}$, $R_{2.0}$, and $R_{2.08}$ show the radius constraints for the cold EOS 
from the combined data analyses in 
Refs.~\cite{Miller21,Raaijmakers21,Pang21,Dittmann24,Rutherford24,Mauviard25,Miller26}.
}
\label{f:eos}
\end{figure*}%...................................................................

\begin{figure*}[t]%..............................................................
\centering
\includegraphics[width=\textwidth]{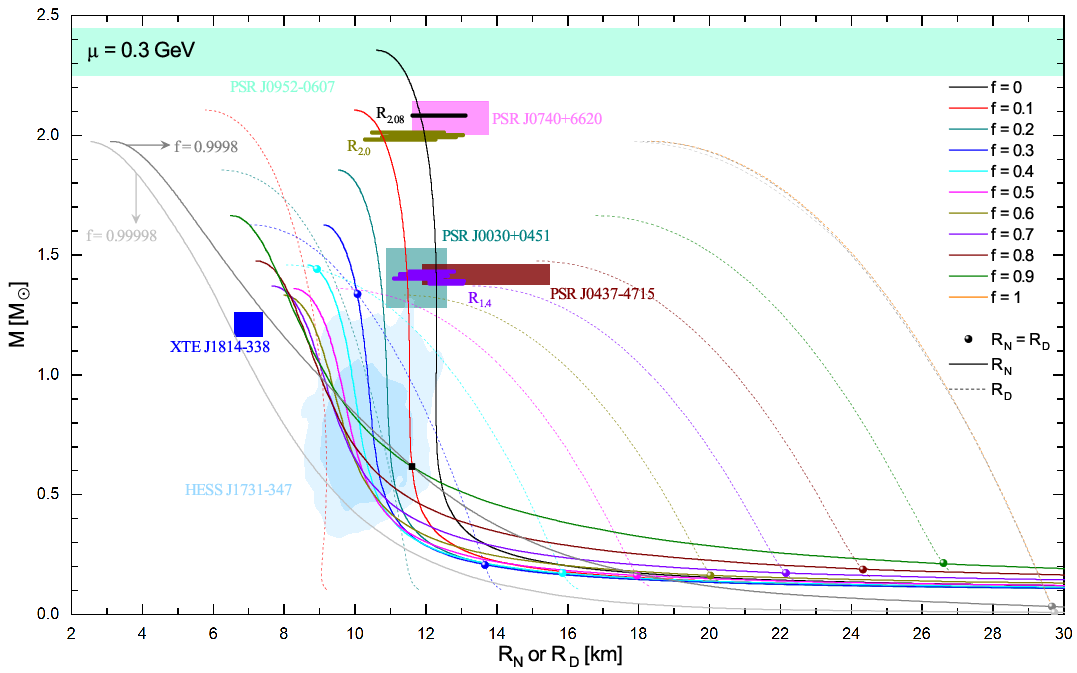}
\caption{
Total gravitational mass $M$ as a function of the optical radius ($R_N$, solid curves) and the dark radius ($R_D$, thin dashed curves) 
of DNSs for DM particle mass $\mu$ = 0.3$\gev$ with different DM fractions $f$. 
The circles mark configurations with $R_N=R_D$.
The shaded color scheme is the same as in Fig.~\ref{f:eos}(b).
}
\label{f:mr}
\end{figure*}%...................................................................

\begin{figure*}[t]%.............................................................
\centering
\includegraphics[width=\textwidth]{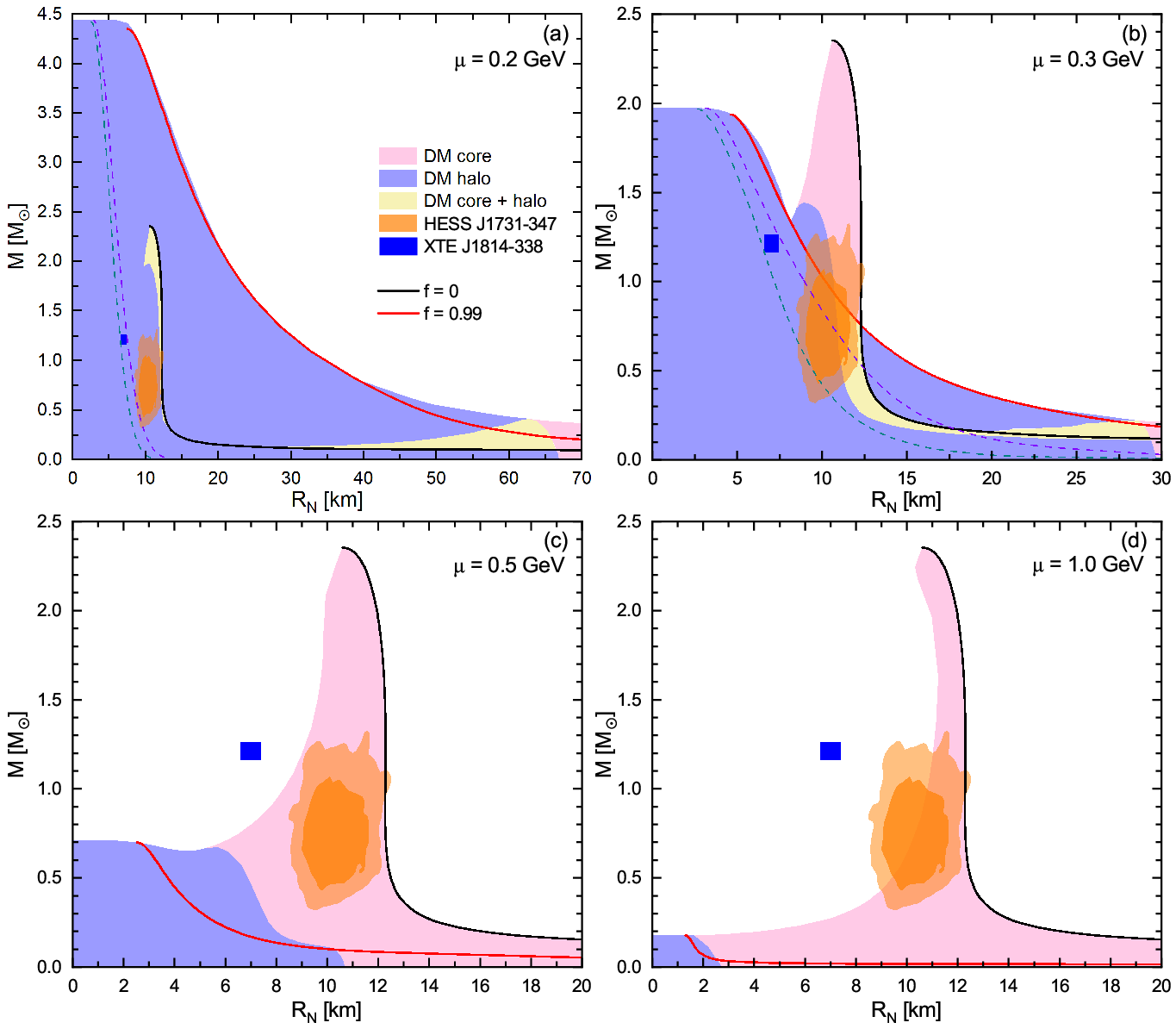}
\caption{The domains of stable DM-core (light pink shading), DM-halo (light purple), 
and both (light yellow) DNSs in the $(M, R_N)$ plane for different DM particle masses. 
The orange and blue shaded regions represent the observational data of HESS J1731-347 and XTE J1814-338, respectively. 
The violet and dark cyan dashed curves
in panels (a) and (b) mark the boundaries of the parameter region compatible with the XTE J1814-338 data. 
The legend identifies the $f=0$ and $f=0.99$ curves.
Note the different axes ranges in the four panels. 
}
\label{f:config}
\end{figure*}%..................................................................

\begin{figure*}[t]%..............................................................
\centering
\includegraphics[width=\textwidth]{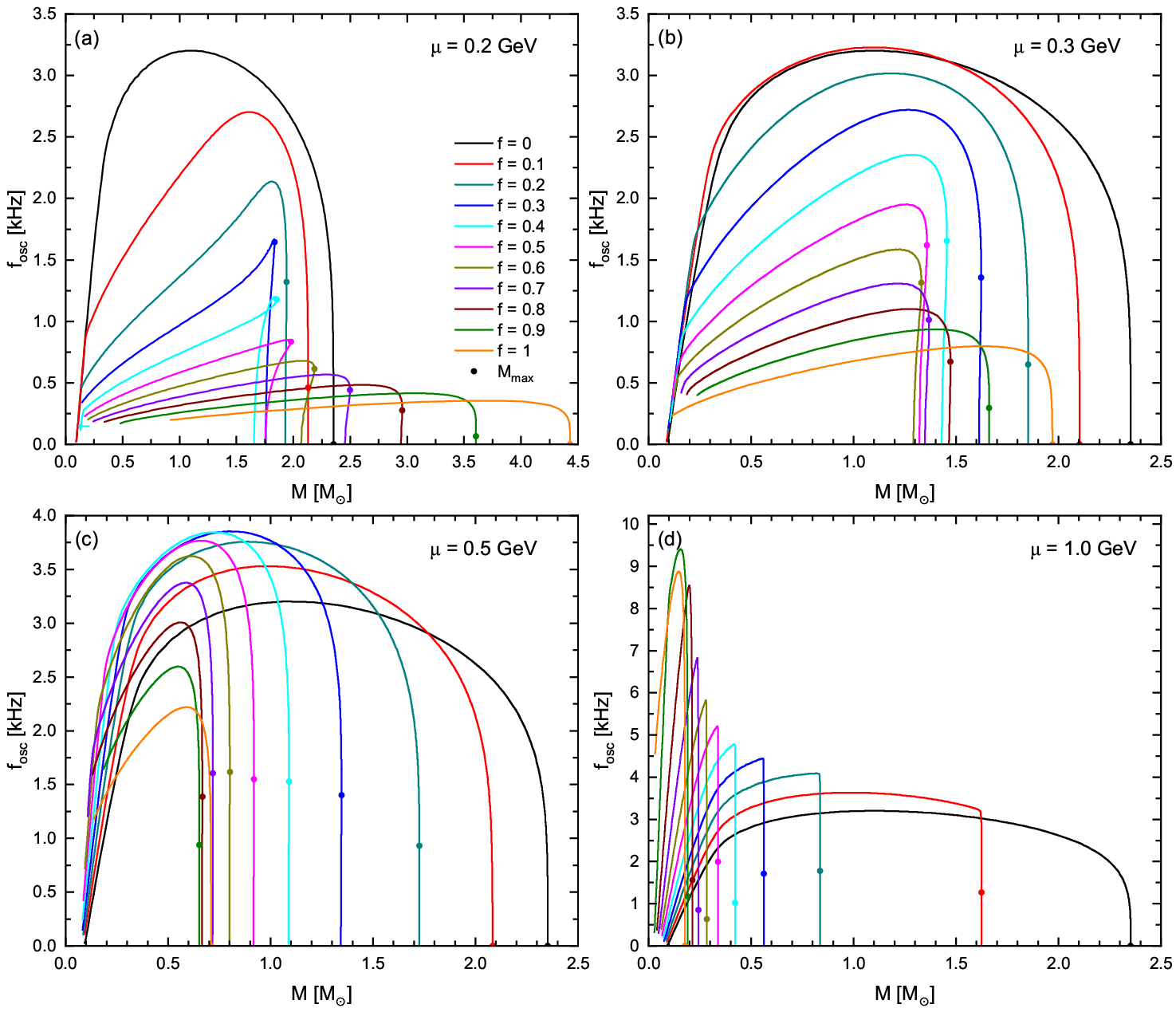}
\caption{Fundamental radial oscillation frequency $f_{\rm{osc}}$ as a function of  the DNS mass $M$ for
DM particle masses $\mu$ = 0.2, 0.3, 0.5, and 1.0 GeV with different DM fractions $f$.
The circles indicate the values of the maximum-mass configurations.}
\label{f:osc}
\end{figure*}%...................................................................

\begin{figure*}[t]%.............................................................
\centering
\includegraphics[width=\textwidth]{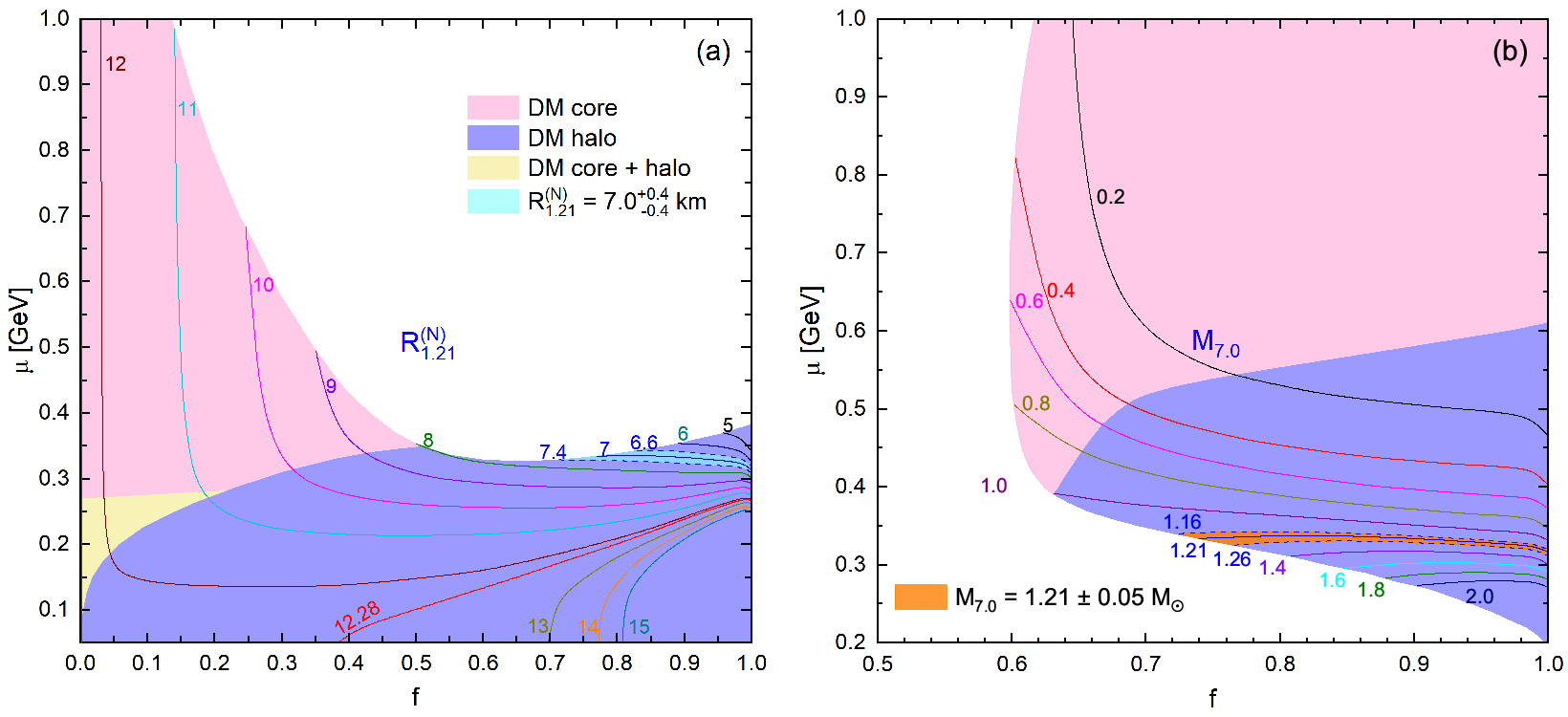}
\caption{
Contour plots of (a) $R^{(N)}_{1.21}$ (in km), the optical radius of a DNS
with total gravitational mass $M=1.21\,M_\odot$, and (b) $M_{7.0}$
(in $M_\odot$), the total gravitational mass of a DNS with optical radius
$R_N=7.0\,\km$, in the $(\mu,f)$ plane.
In panel (a), 12.28 km is the pure NS value.
The light blue region in panel (a) and the orange region in panel (b)
show the parameter ranges compatible with the XTE J1814-338 constraints
$R_N=7.0\pm0.4\,\km$ and $M=1.21\pm0.05\,\ms$, respectively.
The background shading in both panels (light pink, light purple, and light yellow) 
follows the configuration classification in
Fig.~\ref{f:config}.
}
\label{f:r12}
\end{figure*}%..................................................................

\begin{figure}[t]%.............................................................
\centering
\includegraphics[width=\columnwidth]{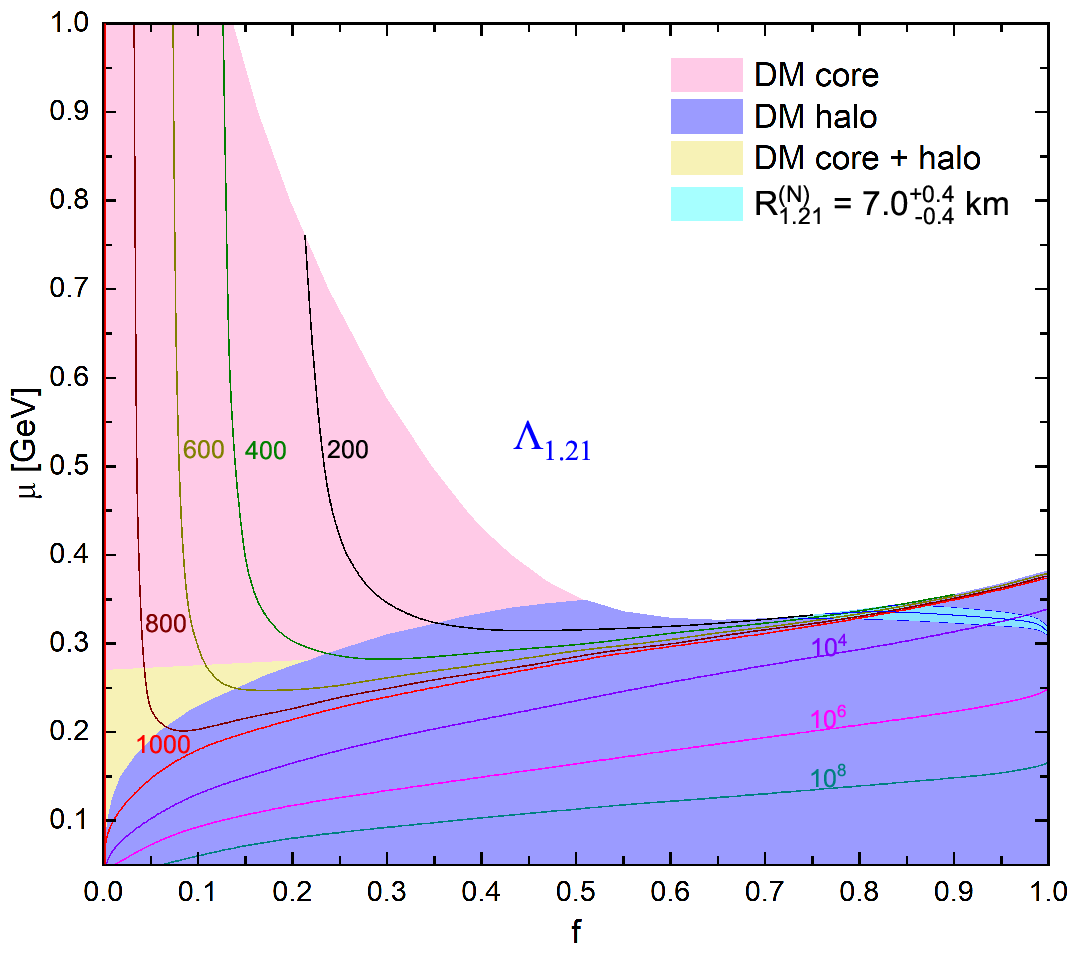}
\caption{Same as Fig.~\ref{f:r12}(a), but for the dimensionless tidal deformability
$\Lambda_{1.21}$ at $M=1.21\,M_\odot$.
The red contour is the pure NS value $\Lambda_{1.21}=1000$; within the
light blue region, the contours give the predicted range of
$\Lambda_{1.21}$ for XTE J1814-338.
}
\label{f:la}
\end{figure}%..................................................................

\begin{figure}[t]%.............................................................
\centering
\includegraphics[width=\columnwidth]{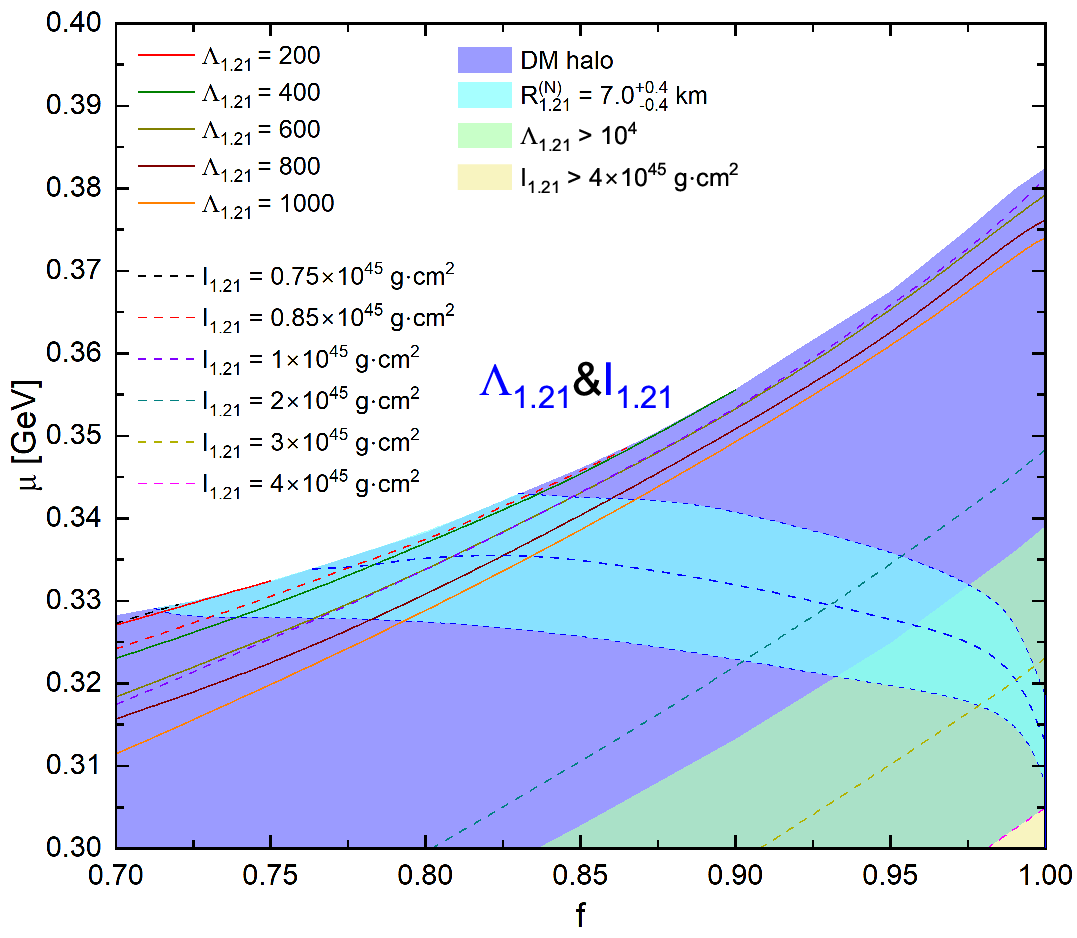}
\caption{
Enlarged contours of the dimensionless tidal deformability
$\Lambda_{1.21}$ (solid curves) and moment of inertia $I_{1.21}$
(dashed curves) for DNSs with $M=1.21\,M_\odot$ in the $(\mu,f)$ plane.
The light green and light yellow regions correspond to
$\Lambda_{1.21}>10^4$ and
$I_{1.21}>4\times10^{45}\,\mathrm{g\cdot cm^2}$, respectively.
The light blue region is the parameter space allowed by the
optical radius constraint of XTE J1814-338.
}
\label{f:pre}
\end{figure}%..................................................................

%-------------------------------------------------------------------------------
\subsection{Equation of state for dark matter}

There are many well-motivated particle candidates for DM, 
which can generally be classified into fermionic and bosonic types \cite{Bertone05,Bertone18}.
Since the fundamental nature of DM remains uncertain, 
we assume in this work that the DM consists of massive self-interacting bosons such as axion-like particles
 \cite{Arvanitaki10,Lee21} in the MeV-GeV range, described by a complex scalar field 
with a quartic self-interaction potential $V(\phi) = \frac{\lambda}{4}|\phi|^4$
\cite{Colpi86,Chavanis12,Karkevandi22}. The corresponding relativistic Lagrangian density is given by
\begin{equation}
\mathcal{L} = \frac{1}{2} \partial_{\alpha} \phi^{*} \partial^{\alpha} \phi - \frac{1}{2} \mu^{2} |\phi|^{2} - \frac{\lambda}{4} |\phi|^{4},
\end{equation}
where $\mu$ is the DM particle mass. 
$\lambda$ is the dimensionless coupling constant that may be related to the DM self-interaction cross section \cite{Collier22,Eby16,XYLi12}.
Assuming that the field forms a Bose-Einstein condensate at sufficiently low temperature, 
the system can be treated in the Thomas-Fermi limit. 
In this regime, the EoS for the self-interacting bosonic DM is
\bal
p_D = \frac{4}{9}\eps_0\left(\sqrt{\frac{3}{4}\frac{\eps_D}{\eps_0} + 1} -1\right)^2,
\label{dmeq:1}
\eal
where $p_D$ and $\eps_D$ represent the DM pressure and the energy density, respectively.
$\varepsilon_0 = {\mu}^4/4\lambda$ is an adjustable parameter. 
Here, we take $\mu$ to be comparable to the nucleon mass, 
as this choice has been shown to produce more noticeable effects on NS observables~\cite{Liu23,Liu24}.
Following Refs.~\cite{Liu23,Liu26}, we adopt $\lambda = \pi$. 
In practice, the exact value is of minor importance, 
as $\mu$ enters the single scale parameter $\varepsilon_0$ with the fourth power. 
A positive $\lambda$ corresponds to the repulsive self-interactions that are essential for stabilizing 
pure bosonic DM stars and preventing gravitational collapse into black holes~\cite{kouvaris13}.
The attractive branch ($\lambda<0$) does not support static bosonic equilibria, 
as such configurations are intrinsically unstable against gravitational collapse \cite{Chavanis11,Eby16}, 
and is therefore not considered here.
 
In the Newtonian limit, 
Eq.~(\ref{dmeq:1}) can be expressed in the following forms:

for low density,
\bal
p_D = \frac{{\eps_D}^2}{16 \eps_0},
\label{dmeq:2}
\eal

for high density,
\bal
p_D = \frac{1}{3}\eps_D.
\label{dmeq:3}
\eal

%-------------------------------------------------------------------------------
\subsection{Hydrostatic configuration}

The properties of DNSs are determined by the Einstein field equations. 
Assuming a static and spherically symmetric spacetime, 
we adopt the standard metric in the form:
\bal
ds^2 = -e^{\nu} dt^2 + \frac{dr^2}{{1 - 2m}/r} + r^2 d\Omega^2,
\eal
where $\nu$ is the metric function.
The equilibrium structure of DNSs is then governed
by solving the two-fluid version of the TOV equations
derived from this metric \cite{Kodama72,Comer99,Sandin09}:
\bal
 \frac{dp_D}{dr} &= -(p_D + \eps_D)\frac{d\nu}{dr}\:,
\\
 \frac{dp_N}{dr} &= -(p_N + \eps_N)\frac{d\nu}{dr}\:,
\label{e:tovn}
\\
 \frac{dm}{dr}   &= 4\pi r^2 \eps\:,
\label{e:tovm}
\\
 \frac{d\nu}{dr} &= \frac{m + 4\pi r^3p}{r(r - 2m)}\:,
\label{e:tovnu}
\eal
where $r$ denotes the radial coordinate. 
$p = p_N+p_D$, $\varepsilon = \varepsilon_N + \varepsilon_D$ 
and $m=m_N+m_D$ are the total pressure,
the total energy density, 
and enclosed mass including contributions from both nuclear and DM components.

The total gravitational mass, 
and the DM fraction inside the DNS are given by
\bal
 M &= m_N(R_N) + m_D(R_D) \:,\\
 \label{eq:gravmass}
f &= \frac{m_D(R_D)}{M} \:,
\eal
where $m_N(R_N)$ and $m_D(R_D)$ are the enclosed masses of nuclear matter and DM, 
while the radii $R_N$ (optical radius) and $R_D$ (dark radius)
are defined by the vanishing of the respective pressures.
The stellar radius $R$ is then defined as the outermost of the two, i.e., $R = \max(R_N, R_D)$.
There are thus in general two scenarios:
DM-core ($R_D<R_N$) or DM-halo ($R_D>R_N$) stars.

For two-fluid DNS configurations, 
two equivalent stability analysis methods are available \cite{Comer99,Leung12,Kain20,Henriques90b,Kain21,Hippert23}. 
The first method is to solve the radial pulsation equations for the fundamental mode frequency: 
configurations become unstable when the (squared) radial oscillation frequency transitions 
from positive to negative \cite{Comer99,Leung12,Kain20}. 
The second method is to identify the critical curve by locating extrema of the total mass 
along fixed-particle-number contours in the two-dimensional parameter space \cite{Henriques90b,Kain21,Hippert23}. 
In this work, 
we adopt the first method and restrict our analysis to configurations that satisfy the stability criterion.
The specific details are presented in Sec.~\ref{Sec.3B}.

%-------------------------------------------------------------------------------
\section{Results}
\label{Sec.3}

%-------------------------------------------------------------------------------
\subsection{Equations of state}

Fig.~\ref{f:eos}(a) illustrates the nucleonic V18 EOS together with the DM EOSs corresponding to various DM particle masses.
In each case, 
the circle marks the central energy density and pressure corresponding to the maximum-mass configuration.
At low and high densities, 
the asymptotic behaviors of the DM EOSs are indicated by the dashed and dotted curves, respectively.
It is evident that smaller DM particle masses lead to lower maximum energy densities and pressures. 
Furthermore, 
all the DM EOSs smoothly connect their respective low-density asymptotic limits (dashed curves) 
and the common high-density limit (dotted curve), consistent with the analytical behaviors 
described by Eqs.~(\ref{dmeq:2}) and (\ref{dmeq:3}). 

The mass-radius relations of pure NS and pure dark stars corresponding to the EOSs 
shown in Fig.~\ref{f:eos}(a) are shown in Fig.~\ref{f:eos}(b).
The maximum NS mass $\mmax$ for the V18 EOS is 2.35$\ms$ and
meets the observational mass constraint from 
the most massive PSR J0952-0607 ($M=2.35\pm0.11\ms$) \cite{Romani25}.
The radii $R_{1.4}=12.28\km$, $R_{2.0}=11.95\km$, and $R_{2.08}=11.82\km$ 
are compatible with the latest mass-radius results of
the NICER mission \cite{Vinciguerra24,Salmi24,Choudhury24}, 
as well as the combined NICER+GW170817 data analyses 
\cite{Miller21,Raaijmakers21,Pang21,Dittmann24,Rutherford24,Mauviard25,Miller26}.
Moreover, 
the mass-radius relation of the pure NS can be compatible with the observational data of HESS J1731-347 \cite{Doroshenko22}, 
whereas the inferred observational data for XTE J1814-338 \cite{Kini24} cannot be satisfied. 
For pure dark stars, 
smaller DM particle masses correspond to larger maximum masses 
and larger minimum radii (i.e., the radii at the maximum mass configurations). 
Especially, 
the mass-radius relation for $\mu \approx 0.3\gev$ is comparable to that of the pure NS, 
beyond (below) this critical value, 
both the mass and the radius will be smaller (larger) than those of pure NSs.

\subsection{DNS mass, radius and configurations}
\label{Sec.3B}

We now study in detail the properties of DNSs within the two-fluid framework. 
Take $\mu$ = 0.3$\gev$ as an example,
Fig.~\ref{f:mr} shows both the mass-optical radius ($M(R_N)$, solid curves) 
and -dark radius ($M(R_D)$, thin dashed curves) relations of DNSs with different DM fractions $f$. 
Each curve terminates at its maximum mass.
The shaded color scheme is same as Fig.~\ref{f:eos}.
One can classify DNSs into two configurations based on the relation between $R_N$ and $R_D$, 
i.e., DM-core and/or DM-halo.
The circles at each fixed $f$ curve denote the transition $R_N$ = $R_D$ between these configurations, if present.

As shown in Fig.~\ref{f:mr}, 
the most prominent effect of DM is the significant modification to the maximum stable mass $\mmax$ 
and the corresponding minimum radius $\rmin$ of NSs with increasing $f$. 
The $\mmax$ gradually decreases with increasing $f$ and begins to deviate from 
the observed mass constraint of PSR J0952-0607 \cite{Romani25} once $f \gtrsim 0.05$.
Subsequently, 
it continues to decrease until $f \approx 0.6$, 
after which it increases again to 1.97$\ms$, 
corr\add{e}sponding to the pure dark star. 
Furthermore, 
attention can be drawn to the variation in the optical radius $R_N$ for a typical NS (e.g., $M=1.4\ms$). 
The $\rn14$ consistently decreases and deviates from the constraints imposed by 
PSR J0030 + 0451 \cite{Vinciguerra24}, PSR J0437-4715 \cite{Miller26}, 
as well as the combined NICER+GW170817 data analyses 
\cite{Miller21,Raaijmakers21,Pang21,Dittmann24,Rutherford24,Mauviard25,Miller26} 
with the successive addition of DM. 
Remarkably, 
when the DM fraction reaches $f = 0.9998$ (dark gray curve) to $f = 0.99998$ (light gray curve), 
the resulting mass-radius relations can simultaneously reproduce the observational data of XTE J1814-338 
(blue shaded region) \cite{Kini24} and HESS J1731-347 (light blue shaded region) \cite{Doroshenko22}. 
Based on this, 
XTE J1814-338 can be interpreted as a DNS with very diluted nuclear matter ($f\rightarrow1$). 
It should be noted that the range of $f$ satisfying the observational data of HESS J1731-347 
and/or XTE J1814-338 varies with $\mu$, 
if present, as discussed in the following detailed analysis.

One can also note the transition behavior between DM-core and DM-halo configurations shown in Fig.~\ref{f:mr}. 
For $f\leq$ 0.2 curves, 
the optical radius $R_N$ remains larger than the dark radius $R_D$ at all stellar masses, 
implying the absence of any intersection point ($R_N = R_D$) and thus purely DM-core configurations. 
As $f$ increases, 
two intersection points emerge for the sequences with $f \lesssim 0.45$, 
indicating that the DNSs undergo the transitions from DM-core to DM-halo and subsequently back again.      
With a further increase in $f$, 
only one intersection point remains, 
corresponding to a monotonic transition from DM-core to the DM-halo along each fixed-$f$ sequence. 
Furthermore, 
the same $(M,R_N)$ may be double- or triple-occupied corresponding to distinct DNS configurations: 
one low-$f$ DM-core star and one/two high-$f $ DM-halo stars, 
as illustrated by the black square in Fig.~\ref{f:mr}.
Such "twin" star are indistinguishable in terms of their mass and optical radius alone, 
yet they exhibit markedly different internal compositions and related properties \cite{Liu23,Liu24,Sun24}.

Given the DM particle mass $\mu$ is still uncertain, 
we therefore present a very rich scenario of DNS configurations in Fig.~\ref{f:config} 
depending on the parameters $\mu$ and $f$. 
The stable domains of DM-core (light pink), DM-halo (light purple), 
and mixed (light yellow) DNSs are displayed in the $(M,R_N)$ plane for $\mu$ = 0.2, 0.3, 0.5, 1.0$\gev$.
As shown in the figure, when $\mu$ increases from 0.2 to 1.0$\gev$, 
the overall configuration landscape evolves markedly.
For larger $\mu$ values (0.5 and 1.0$\gev$), 
the maximum stable mass predominantly decreases as DM is added, 
and this effect becomes more pronounced with increasing $\mu$. 
In this regime, the DM-core configurations become dominant, 
whereas the DM-halo configurations are confined to domains with extremely large $f$. 
As $\mu$ increases further ($\mu > 1.0\gev$), nearly all stellar configurations become DM-core. 

For small $\mu$ values (0.2 and 0.3$\gev$), 
the maximum stable mass first decreases with increasing $f$, 
then increases again and for $\mu=0.2\gev$ it even exceeds that of a pure NS. 
The configuration plot becomes more complex: 
(a) a mixed region (light yellow) containing both DM-core 
and DM-halo configurations emerges because certain $(M, R_N)$ points are double- or triple-occupied, 
as discussed in Fig.~\ref{f:mr}; 
and (b) DM-halo configurations with large $f$ values become increasingly significant 
and occupying progressively larger regions.
With the decreasing $\mu$, 
the optical radius $R_N$ for a typical NS varies between 0 ($f=1$) 
and values exceeding the pure NS radius $(f=0)$.

For $\mu$ = 0.2$\gev$, XTE J1814-338 can be naturally interpreted as a DM-halo star with high $f$
ranging from 0.999999982 to 0.9999999935 (corresponding to the violet and dark cyan dashed curves in Fig.~\ref{f:config}(a)), 
which is consistent with the conclusion shown in Fig.~\ref{f:mr} for $\mu$ = 0.3$\gev$. 
However, the allowable range of $f$ for XTE J1814-338
(i.e., the region between the violet and orange dashed curves) only partially overlaps 
with the observational data from HESS J1731-347.
For $\mu$ = 0.5 and 1.0$\gev$, 
XTE J1814-338 can never be interpreted as a DM admixture for any value of the DM mass fraction. 
This implies the existence of a threshold DM particle mass, 
above which XTE J1814-338 can no longer be explained as a DM-admixed configuration.

Regarding the stability of the DNS configurations shown in 
Figs.~\ref{f:mr} and \ref{f:config} against radial oscillation perturbations, 
Fig.~\ref{f:osc} shows the radial oscillation frequencies $f_{\rm osc}$ as functions of DNS masses 
$M$ for $\mu = 0.2, 0.3, 0.5, 1.0\gev$ with various $f$. 
In all cases the $f_{\rm osc}$ first rises and then drops with increasing $M$. 
For $\mu$ = 0.5 and 1.0 GeV it vanishes essentially at the $M_{\text{max}}$ configuration, 
while for smaller $\mu$ it remains positive beyond the $M_{\text{max}}$ 
and vanishes further down the descending branch, 
consistent with the trend reported in Ref.~\cite{Kain20}. 
All configurations considered in this work lie on the stable branch with $f_{\rm osc} > 0$.

Following the preceding discussion,
we can thus explore the allowed parameter space of XTE J1814-338 in the ($\mu, f$) plane. 
Fig.~\ref{f:r12}(a) shows the contour plot of the optical radius $R^{(N)}_{1.21}$ 
(at $M=1.21\ms$, the central mass of XTE J1814-338). 
Only DM-core configurations (light pink domain) appear for $\mu\gtrsim 0.35\gev$, 
and $R^{(N)}_{1.21}$ decreases slightly with increasing $f$ due to the gravitational pull of DM core 
until the stability boundary at $M = 1.21\ms$ is reached. 
For smaller $\mu$, a broader range of $f$ is permitted, 
and $R^{(N)}_{1.21}$ can decrease to approximately $5\km$ for $\mu = 0.35\gev$.
For $\mu< 0.35\gev$, 
DM-halo stars (light purple domain) dominate and $R^{(N)}_{1.21}$ varies significantly without an upper limit on $f$. 
We can roughly divide it into two cases: $0.265<\mu< 0.35\gev$ and $\mu< 0.265\gev$. 
In the first case, $R^{(N)}_{1.21}$ can decrease to arbitrarily small values for very large $f$, 
whereas in the second case $R^{(N)}_{1.21}$ begins to increase again and can even exceed that of the pure NS 
(12.28$\km$, red dashed curve) due to the ``stretching" effect, 
where the DM halo exerts an outward gravitational pull on the nuclear matter \cite{Ellis18,Ivanytskyi20,Karkevandi22}. 

Note that the optical radius tends to zero in the limit $f\rightarrow1$, 
corresponding to a dark star.
We therefore delineate the full $(\mu, f)$ parameter space that 
satisfies the optical radius constraint $R_N = 7.0 \pm 0.4\km$ of XTE J1814-338, 
shown as the light blue shaded region in Fig.~\ref{f:r12}.
It is evident that XTE J1814-338 can only be interpreted as a DM-halo star. 
A relatively broad parameter range is found within $0.31\gev \lesssim \mu \lesssim 0.345\gev$ and $0.7 \lesssim f \lesssim 1$. 
With decreasing $\mu$, 
dark stars containing only a minute fraction of nuclear matter can still reproduce the observed radius of XTE J1814-338. 
If future observations confirm its DM-halo nature, an upper limit can then be placed on the DM particle mass.
We emphasize that a similar analysis in the $(\mu, f)$ plane was performed for $M = 1.4\ms$ in Refs.~\cite{Liu23,Liu24}. 
The present work extends this to $M = 1.21M_\odot$, the mass for XTE J1814-338, 
yielding quantitatively different bounds on the allowed $\mu$ and $f$ ranges.

To test the sensitivity to the observational mass uncertainty,
Fig.~\ref{f:r12}(b) shows the contour plot of the total DNS mass $M_{7.0}$ 
(at $R_N=7.0\km$, the central radius of XTE J1814-338). 
As demonstrated in Fig.~\ref{f:r12}(b), 
the allowed parameter space remains qualitatively unchanged across the entire mass range 
($1.16\ms\leq M_{7.0}\leq 1.26\ms$, orange shaded region):
only DM-halo configurations with $f \to 1$ and $\mu \lesssim 0.345\gev$ can reproduce the observed compactness. 
The upper bound on $\mu$ shifts only slightly with $M$, 
confirming the robustness of our conclusions against the mass uncertainty.

%-------------------------------------------------------------------------------
\subsection{DNS tidal deformability and moment of inertia}
\label{Sec.3C}
 
We then examine the effect of DM on the observable quantity most directly related to the stellar radius, 
namely the dimensionless tidal deformability $\la$, 
which is obtained from the tidal Love number $k_2$ \cite{Hartle67,Binnington09,Damour10}:
\be
 \la = \frac23 \bigg(\frac{R}{M}\bigg)^5 k_2 \:,
\label{eq:la}
\ee
where $R={\rm max}(R_N,R_D)$ denotes the stellar radius, i.e., 
the outermost DNS radius in our formalism. 
The equations required to compute $k_2$ within the two-fluid formalism are detailed, 
for instance, in Refs.~\cite{Leung22,Karkevandi22,Zhang23}.
The $\la$ is extracted from the phase evolution of the 
gravitational-wave signal during the inspiral of a binary NS merger.
The leading-order tidal effect enters at the fifth post-Newtonian order, 
and the combined observable is an effective tidal deformability $\tilde{\Lambda}$, 
a mass-weighted combination of the individual $\Lambda_1$ and $\Lambda_2$ \cite{Flanagan08,Hinderer08,Abbott18}. 
For GW170817, the LIGO-Virgo Collaboration constrained $70\leq\l14\leq 580$. 
Future detectors will further improve this sensitivity \cite{Abbott18}.

Fig.~\ref{f:la} presents the contour plot of dimensionless tidal deformability at $M=1.21\ms$, 
denoted as $\la_{1.21}$, 
in the $(\mu, f)$ plane, 
analogous to $R^{(N)}_{1.21}$ in Fig.~\ref{f:r12}. 
As indicated by Eq.~(\ref{eq:la}), 
$\la_{1.21}$ is highly sensitive to the fifth power of the $R_{1.21}$, 
which allows its overall behavior to be readily understood.
As analyzed in Sec.~\ref{Sec.3B}, 
for $\mu\gtrsim 0.35\gev$ only DM-core stars exist (light pink domain). 
In this regime, the variation of $\la_{1.21}$ with $f$ follows the same trend as that of $R^{(N)}_{1.21}$; 
$\la_{1.21}$ decreases from the pure NS value 
$\la_{1.21}=1000$ (red contour) to below 100, by about an order of magnitude.
As $\mu$ decreases, 
DM-halo configurations (light purple region) become dominant, 
and $\la_{1.21}$ becomes strongly dependent on the dark radius $R^{(D)}_{1.21}$. 
Even relatively small DM fractions can result in extremely large $\la_{1.21}$ values with $\mu\rightarrow0$ (light yellow domain). 
Combining this trend with the allowed ($\mu, f$) parameter space inferred from Fig.~\ref{f:r12} (light blue shaded region), 
the $\la$ of XTE J1814-338 can be theoretically constrained, 
increasing from $\sim 200$ to extremely large values as $\mu$ decreases (e.g., reaching
$\sim 10^{6}$ for $\mu = 0.25\gev$).
The behavior of $\la_{1.21}$ in the $(\mu, f)$ plane is qualitatively similar to 
that reported in Refs.~\cite{Liu23,Liu24} for $\la_{1.4}$. 
The mass shift from $1.4\ms$ to $1.21\ms$ primarily affects the numerical values of the tidal deformability, 
particularly in the DM-halo regime.

For clarity, 
we further magnify the contour plot around the allowed $(\mu, f)$ parameter space 
(light blue shaded region) in Fig.~\ref{f:la}  and present it in Fig.~\ref{f:pre}. 
In addition, 
we examine another observable quantity that characterizes the rotational properties of NSs, 
namely the moment of inertia $I$, 
defined as
\be
I = \frac{J}{\Omega},
\label{eq:MOI}
\ee
where $J$ is the angular momentum and $\Omega\equiv2\pi f$ is 
the angular frequency measured by a distant observer (i.e., the pulsar rotation frequency).

As reported in Refs.~\cite{Lyne04,Lattimer05,Kramer09}, 
the moment of inertia can be inferred from timing observations of binary radio pulsar systems,
 thereby providing crucial constraints on the equation of state at high densities. 
In the present work, 
$I$ is computed within the slow-rotation framework of Hartle and Thorne \cite{Hartle67,Hartle68}. 
We solve the first-order differential equation for the auxiliary function $w(r)$,
\begin{equation}
\frac{dw}{dr} = \frac{4\pi(\epsilon + p)(4 + w) r}{1 - 2m/r}
- \frac{w(3 + w)}{r},
\end{equation}
simultaneously with the two-fluid TOV equations. 
At the stellar surface $r = R$,
the moment of inertia is then obtained as
\begin{equation}
I = \frac{w(R) R^3}{6 + 2w(R)}.
\end{equation}

Fig.~\ref{f:pre} also presents the contour plot of $I$ at $M = 1.21\ms$, 
denoted as $I_{1.21}$ (dotted curves), in the ($\mu, f$) plane.
For a given $f$, 
the $I$ increases with decreasing $\mu$ 
owing to its quadratic dependence on the stellar radius.
For a fixed $\mu$, 
the progressive addition of DM further expands the stellar radius 
and consequently results in a higher $I$.
We can therefore estimate the predicted values of $\la$ and $I$ 
for XTE J1814-338 as a DNS from Fig.~\ref{f:pre}. 
Only lower bounds can be established for these quantities.
As discussed above, as $\mu$ decreases, the allowed parameter region
narrows toward $f\to1$, while $R^{(D)}_{1.21}$ increases and approaches
the pure DM star radius at $M=1.21\,M_\odot$.
In the DM halo regime, stellar radius $R=R_D>R_N$.
Since both $\Lambda$ and $I$
increase with $R$, 
and there are no direct observational constraints on $\mu$ or $R_D$, 
these quantities have no upper bound. 
The resulting limits are $\Lambda \gtrsim 200$ (red solid curve)
and $I \gtrsim 0.75\times10^{45}~\mathrm{g\cdot cm^{2}}$ (black dashed curve),
respectively.

Note that Figs.~\ref{f:la} and \ref{f:pre} present the tidal deformability 
and moment of inertia evaluated at the central mass value $M = 1.21\ms$. 
The corresponding results at fixed radius $R_N = 7.0\km$ are not shown, 
but are expected to be close to those presented here,
since the allowed $(\mu, f)$ region is not sensitive to the mass variation within $\pm 0.05\ms$.

Future high-precision measurements from binary radio pulsars 
or gravitational-wave observations could therefore target XTE J1814-338 
and provide valuable constraints on the properties of DM particles.

%-------------------------------------------------------------------------------
\section{Summary}
\label{Sec.4}

Combining a realistic and well-constrained BHF EOS of pure nuclear matter with a generic DM EOS of interacting bosons, 
we have investigated the possibility that the ultracompact object XTE J1814-338 is a DNS.
This analysis covers a wide parameter space, 
in which the DM particle mass $\mu$ varies on the scale of nucleon masses and DM fraction $f$ can reach up to 100\%, 
corresponding to pure dark stars. 
This allows to comprehensively scan the quantitative influence of DM on the properties of NSs 
including masses, radii and stellar configurations.

We found that for large $\mu$ values, 
both the radii and the maximum masses of DNSs gradually decrease with added DM, 
and DM-core stars become dominant. 
With decreasing $\mu$, the fraction of DM-core stars diminishes, 
while the maximum masses and radii only decrease slightly. 
In contrast, 
DM-halo stars can reach extremely large masses and radii as $\mu\rightarrow0$.
Based on the inferred observational data from XTE J1814-338, 
we constructed the contour plot of the optical radius at $M$ = 1.21$\ms$ in the ($\mu,f$) plane. 
The allowed $(\mu,f)$ parameter space indicates that XTE J1814-338 can only be a DM-halo star with high $f$, 
qualitatively consistent with the conclusions of Ref.~\cite{Pitz25} 
using a substantially stiffer DM EOS, although the detailed microphysical assumptions differ. 
Once $\mu$ exceeds 0.345$\gev$, 
XTE J1814-338 can never be interpreted as a DM admixture.
However, 
the formation of such DM-rich objects would require exotic capture or formation mechanisms 
\cite{Goldman13,Kouvaris15,DiGiovanni20}.
If the DM-halo nature of XTE J1814-338 is confirmed by subsequent observations, 
an observational upper limit on the DM particle mass could be established.

Furthermore, 
we theoretically predicted the values of two other observable quantities of XTE J1814-338:
dimensionless tidal deformability $\la$ and moment of inertia $I$. 
Both quantities scale strongly with the stellar radius, 
and their behaviors across the $(\mu,f)$ parameter space 
can thus be inferred from the previous radius analysis.
Our results indicated that, 
if XTE J1814-338 is interpreted as a DNS, 
these observable quantities have theoretical lower limits of 
$\la \gtrsim 200$ and $I \gtrsim 0.75\times10^{45}~\mathrm{g\cdot cm^{2}}$, respectively. 
These theoretical lower limits may serve as potential observational constraints 
for future missions aiming to distinguish DM-admixed from ordinary compact stars.

%\vspace{3mm}
%-------------------------------------------------------------------------------
\section*{Acknowledgments}
We would like to thank Dr. Jin-Biao Wei for insightful discussions.
This work is sponsored by the National Key R\&D Program of of China No.~2022YFA1602303, 
the National Natural Science Foundation of China under Grant Nos.~12575134, 11975132 and 12205158, 
the Shandong Provincial Natural Science Foundation, China ZR2025QC1487, ZR2021QA037 and ZR2022JQ04, 
and the Qingdao Natural Science Foundation 25-1-1-4-zyyd-jch.

%-------------------------------------------------------------------------------
\def\aap{A\&A}
\def\apjl{Astrophys. J. Lett.}
\def\araa{Annu. Rev. Astron. Astrophys.}
\def\epja{Eur. Phys. J. A}
\def\epjc{Eur. Phys. J. C}
\def\jcap{J. Cosmol. Astropart. Phys.}
\def\jpg{J. Phys. G}
\def\mnras{Mon. Not. R. Astron. Soc.}
\def\npb{Nucl. Phys. B}
\def\physrep{Phys. Rep.}
\def\plb{Phys. Lett. B}
\def\ppnp{Prog. Part. Nucl. Phys.}
\def\rpp{Rep. Prog. Phys.}
\def\cpc{Chin. Phys. C}
\bibliographystyle{apsrev4-1}
%\nocite{*}
\bibliography{dmf}

\begin{thebibliography}{106}%
\makeatletter
\providecommand \@ifxundefined [1]{%
 \@ifx{#1\undefined}
}%
\providecommand \@ifnum [1]{%
 \ifnum #1\expandafter \@firstoftwo
 \else \expandafter \@secondoftwo
 \fi
}%
\providecommand \@ifx [1]{%
 \ifx #1\expandafter \@firstoftwo
 \else \expandafter \@secondoftwo
 \fi
}%
\providecommand \natexlab [1]{#1}%
\providecommand \enquote  [1]{``#1''}%
\providecommand \bibnamefont  [1]{#1}%
\providecommand \bibfnamefont [1]{#1}%
\providecommand \citenamefont [1]{#1}%
\providecommand \href@noop [0]{\@secondoftwo}%
\providecommand \href [0]{\begingroup \@sanitize@url \@href}%
\providecommand \@href[1]{\@@startlink{#1}\@@href}%
\providecommand \@@href[1]{\endgroup#1\@@endlink}%
\providecommand \@sanitize@url [0]{\catcode `\\12\catcode `\$12\catcode
  `\&12\catcode `\#12\catcode `\^12\catcode `\_12\catcode `\%12\relax}%
\providecommand \@@startlink[1]{}%
\providecommand \@@endlink[0]{}%
\providecommand \url  [0]{\begingroup\@sanitize@url \@url }%
\providecommand \@url [1]{\endgroup\@href {#1}{\urlprefix }}%
\providecommand \urlprefix  [0]{URL }%
\providecommand \Eprint [0]{\href }%
\providecommand \doibase [0]{http://dx.doi.org/}%
\providecommand \selectlanguage [0]{\@gobble}%
\providecommand \bibinfo  [0]{\@secondoftwo}%
\providecommand \bibfield  [0]{\@secondoftwo}%
\providecommand \translation [1]{[#1]}%
\providecommand \BibitemOpen [0]{}%
\providecommand \bibitemStop [0]{}%
\providecommand \bibitemNoStop [0]{.\EOS\space}%
\providecommand \EOS [0]{\spacefactor3000\relax}%
\providecommand \BibitemShut  [1]{\csname bibitem#1\endcsname}%
\let\auto@bib@innerbib\@empty
%</preamble>
\bibitem [{\citenamefont {Steiner}\ \emph {et~al.}(2005)\citenamefont
  {Steiner}, \citenamefont {Prakash}, \citenamefont {Lattimer},\ and\
  \citenamefont {Ellis}}]{Steiner05}%
  \BibitemOpen
  \bibfield  {author} {\bibinfo {author} {\bibfnamefont {A.}~\bibnamefont
  {Steiner}}, \bibinfo {author} {\bibfnamefont {M.}~\bibnamefont {Prakash}},
  \bibinfo {author} {\bibfnamefont {J.}~\bibnamefont {Lattimer}}, \ and\
  \bibinfo {author} {\bibfnamefont {P.}~\bibnamefont {Ellis}},\ }\href
  {\doibase https://doi.org/10.1016/j.physrep.2005.02.004} {\bibfield
  {journal} {\bibinfo  {journal} {\physrep}\ }\textbf {\bibinfo {volume}
  {411}},\ \bibinfo {pages} {325} (\bibinfo {year} {2005})}\BibitemShut
  {NoStop}%
\bibitem [{\citenamefont {Burgio}\ \emph {et~al.}(2021)\citenamefont {Burgio},
  \citenamefont {Schulze}, \citenamefont {Vida{\~n}a},\ and\ \citenamefont
  {Wei}}]{Burgio21}%
  \BibitemOpen
  \bibfield  {author} {\bibinfo {author} {\bibfnamefont {G.~F.}\ \bibnamefont
  {Burgio}}, \bibinfo {author} {\bibfnamefont {H.-J.}\ \bibnamefont {Schulze}},
  \bibinfo {author} {\bibfnamefont {I.}~\bibnamefont {Vida{\~n}a}}, \ and\
  \bibinfo {author} {\bibfnamefont {J.-B.}\ \bibnamefont {Wei}},\ }\href
  {\doibase https://doi.org/10.1016/j.ppnp.2021.103879} {\bibfield  {journal}
  {\bibinfo  {journal} {\ppnp}\ }\textbf {\bibinfo {volume} {120}},\ \bibinfo
  {pages} {103879} (\bibinfo {year} {2021})}\BibitemShut {NoStop}%
\bibitem [{\citenamefont {Abbott}\ \emph {et~al.}(2017)\citenamefont {Abbott}
  \emph {et~al.}}]{Abbott17}%
  \BibitemOpen
  \bibfield  {author} {\bibinfo {author} {\bibfnamefont {B.~P.}\ \bibnamefont
  {Abbott}} \emph {et~al.} (\bibinfo {collaboration} {LIGO Scientific
  Collaboration and Virgo Collaboration}),\ }\href {\doibase
  10.1103/PhysRevLett.119.161101} {\bibfield  {journal} {\bibinfo  {journal}
  {Phys. Rev. Lett.}\ }\textbf {\bibinfo {volume} {119}},\ \bibinfo {pages}
  {161101} (\bibinfo {year} {2017})}\BibitemShut {NoStop}%
\bibitem [{\citenamefont {Abbott}\ \emph {et~al.}(2018)\citenamefont {Abbott}
  \emph {et~al.}}]{Abbott18}%
  \BibitemOpen
  \bibfield  {author} {\bibinfo {author} {\bibfnamefont {B.~P.}\ \bibnamefont
  {Abbott}} \emph {et~al.},\ }\href {\doibase 10.1103/PhysRevLett.121.161101}
  {\bibfield  {journal} {\bibinfo  {journal} {Phys. Rev. Lett.}\ }\textbf
  {\bibinfo {volume} {121}},\ \bibinfo {eid} {161101} (\bibinfo {year}
  {2018})}\BibitemShut {NoStop}%
\bibitem [{\citenamefont {{\it et al.}}(2020)}]{Cromartie20}%
  \BibitemOpen
  \bibfield  {author} {\bibinfo {author} {\bibfnamefont {H.~T.~C.}\
  \bibnamefont {{\it et al.}}},\ }\href {\doibase 10.1038/s41550-019-0880-2}
  {\bibfield  {journal} {\bibinfo  {journal} {Nat. Astron.}\ }\textbf {\bibinfo
  {volume} {4}},\ \bibinfo {pages} {72} (\bibinfo {year} {2020})}\BibitemShut
  {NoStop}%
\bibitem [{\citenamefont {Miller}\ \emph {et~al.}(2019)\citenamefont {Miller}
  \emph {et~al.}}]{Miller19}%
  \BibitemOpen
  \bibfield  {author} {\bibinfo {author} {\bibfnamefont {M.~C.}\ \bibnamefont
  {Miller}} \emph {et~al.},\ }\href {\doibase 10.3847/2041-8213/ab50c5}
  {\bibfield  {journal} {\bibinfo  {journal} {Astrophys. J.}\ }\textbf
  {\bibinfo {volume} {887}},\ \bibinfo {pages} {L24} (\bibinfo {year}
  {2019})}\BibitemShut {NoStop}%
\bibitem [{\citenamefont {Riley}\ \emph {et~al.}(2019)\citenamefont {Riley}
  \emph {et~al.}}]{Riley19}%
  \BibitemOpen
  \bibfield  {author} {\bibinfo {author} {\bibfnamefont {T.~E.}\ \bibnamefont
  {Riley}} \emph {et~al.},\ }\href {\doibase 10.3847/2041-8213/ab481c}
  {\bibfield  {journal} {\bibinfo  {journal} {Astrophys. J.}\ }\textbf
  {\bibinfo {volume} {887}},\ \bibinfo {pages} {L21} (\bibinfo {year}
  {2019})}\BibitemShut {NoStop}%
\bibitem [{\citenamefont {Riley}\ \emph {et~al.}(2021)\citenamefont {Riley}
  \emph {et~al.}}]{Riley21}%
  \BibitemOpen
  \bibfield  {author} {\bibinfo {author} {\bibfnamefont {T.~E.}\ \bibnamefont
  {Riley}} \emph {et~al.},\ }\href {\doibase 10.3847/2041-8213/ac0a81}
  {\bibfield  {journal} {\bibinfo  {journal} {Astrophys. J. Lett.}\ }\textbf
  {\bibinfo {volume} {918}},\ \bibinfo {pages} {L27} (\bibinfo {year}
  {2021})}\BibitemShut {NoStop}%
\bibitem [{\citenamefont {Miller}\ \emph {et~al.}(2021)\citenamefont {Miller}
  \emph {et~al.}}]{Miller21}%
  \BibitemOpen
  \bibfield  {author} {\bibinfo {author} {\bibfnamefont {M.~C.}\ \bibnamefont
  {Miller}} \emph {et~al.},\ }\href {\doibase 10.3847/2041-8213/ac089b}
  {\bibfield  {journal} {\bibinfo  {journal} {Astrophys. J. Lett.}\ }\textbf
  {\bibinfo {volume} {918}},\ \bibinfo {pages} {L28} (\bibinfo {year}
  {2021})}\BibitemShut {NoStop}%
\bibitem [{\citenamefont {Choudhury}\ \emph {et~al.}(2024)\citenamefont
  {Choudhury} \emph {et~al.}}]{Choudhury24}%
  \BibitemOpen
  \bibfield  {author} {\bibinfo {author} {\bibfnamefont {D.}~\bibnamefont
  {Choudhury}} \emph {et~al.},\ }\href {\doibase 10.3847/2041-8213/ad5a6f}
  {\bibfield  {journal} {\bibinfo  {journal} {Astrophys. J. Lett.}\ }\textbf
  {\bibinfo {volume} {971}},\ \bibinfo {pages} {L20} (\bibinfo {year}
  {2024})}\BibitemShut {NoStop}%
\bibitem [{\citenamefont {Romani}\ \emph {et~al.}(2022)\citenamefont {Romani},
  \citenamefont {Kandel}, \citenamefont {Filippenko}, \citenamefont {Brink},\
  and\ \citenamefont {Zheng}}]{Romani22}%
  \BibitemOpen
  \bibfield  {author} {\bibinfo {author} {\bibfnamefont {R.~W.}\ \bibnamefont
  {Romani}}, \bibinfo {author} {\bibfnamefont {D.}~\bibnamefont {Kandel}},
  \bibinfo {author} {\bibfnamefont {A.~V.}\ \bibnamefont {Filippenko}},
  \bibinfo {author} {\bibfnamefont {T.~G.}\ \bibnamefont {Brink}}, \ and\
  \bibinfo {author} {\bibfnamefont {W.}~\bibnamefont {Zheng}},\ }\href
  {\doibase 10.3847/2041-8213/ac8007} {\bibfield  {journal} {\bibinfo
  {journal} {\apjl}\ }\textbf {\bibinfo {volume} {934}},\ \bibinfo {pages}
  {L17} (\bibinfo {year} {2022})}\BibitemShut {NoStop}%
\bibitem [{\citenamefont {Doroshenko}\ \emph {et~al.}(2022)\citenamefont
  {Doroshenko}, \citenamefont {Suleimanov}, \citenamefont {P{\"u}hlhofer},\
  and\ \citenamefont {Santangelo}}]{Doroshenko22}%
  \BibitemOpen
  \bibfield  {author} {\bibinfo {author} {\bibfnamefont {V.}~\bibnamefont
  {Doroshenko}}, \bibinfo {author} {\bibfnamefont {V.}~\bibnamefont
  {Suleimanov}}, \bibinfo {author} {\bibfnamefont {G.}~\bibnamefont
  {P{\"u}hlhofer}}, \ and\ \bibinfo {author} {\bibfnamefont {A.}~\bibnamefont
  {Santangelo}},\ }\href {\doibase 10.1038/s41550-022-01800-1} {\bibfield
  {journal} {\bibinfo  {journal} {Nat. Astron.}\ }\textbf {\bibinfo {volume}
  {6}},\ \bibinfo {pages} {1444} (\bibinfo {year} {2022})}\BibitemShut
  {NoStop}%
\bibitem [{\citenamefont {Abac}\ \emph {et~al.}(2024)\citenamefont {Abac} \emph
  {et~al.}}]{Abac24}%
  \BibitemOpen
  \bibfield  {author} {\bibinfo {author} {\bibfnamefont {A.~G.}\ \bibnamefont
  {Abac}} \emph {et~al.} (\bibinfo {collaboration} {The LIGO Scientific
  Collaboration, the Virgo Collaboration, and the KAGRA Collaboration}),\
  }\href {\doibase 10.3847/2041-8213/ad5beb} {\bibfield  {journal} {\bibinfo
  {journal} {\apjl}\ }\textbf {\bibinfo {volume} {970}},\ \bibinfo {pages}
  {L34} (\bibinfo {year} {2024})}\BibitemShut {NoStop}%
\bibitem [{\citenamefont {Barr}\ \emph {et~al.}(2024)\citenamefont {Barr} \emph
  {et~al.}}]{Ewan24}%
  \BibitemOpen
  \bibfield  {author} {\bibinfo {author} {\bibfnamefont {E.~D.}\ \bibnamefont
  {Barr}} \emph {et~al.},\ }\href {\doibase 10.1126/science.adg3005} {\bibfield
   {journal} {\bibinfo  {journal} {Science}\ }\textbf {\bibinfo {volume}
  {383}},\ \bibinfo {pages} {275} (\bibinfo {year} {2024})}\BibitemShut
  {NoStop}%
\bibitem [{\citenamefont {Routaray}\ \emph
  {et~al.}(2023{\natexlab{a}})\citenamefont {Routaray}, \citenamefont
  {Mohanty}, \citenamefont {Das}, \citenamefont {Ghosh}, \citenamefont
  {Kalita}, \citenamefont {Parmar},\ and\ \citenamefont {Kumar}}]{Routaray23}%
  \BibitemOpen
  \bibfield  {author} {\bibinfo {author} {\bibfnamefont {P.}~\bibnamefont
  {Routaray}}, \bibinfo {author} {\bibfnamefont {S.~R.}\ \bibnamefont
  {Mohanty}}, \bibinfo {author} {\bibfnamefont {H.}~\bibnamefont {Das}},
  \bibinfo {author} {\bibfnamefont {S.}~\bibnamefont {Ghosh}}, \bibinfo
  {author} {\bibfnamefont {P.}~\bibnamefont {Kalita}}, \bibinfo {author}
  {\bibfnamefont {V.}~\bibnamefont {Parmar}}, \ and\ \bibinfo {author}
  {\bibfnamefont {B.}~\bibnamefont {Kumar}},\ }\href {\doibase
  10.1088/1475-7516/2023/10/073} {\bibfield  {journal} {\bibinfo  {journal}
  {\jcap}\ }\textbf {\bibinfo {volume} {2023}},\ \bibinfo {pages} {073}
  (\bibinfo {year} {2023}{\natexlab{a}})}\BibitemShut {NoStop}%
\bibitem [{\citenamefont {Hong}\ and\ \citenamefont {Ren}(2024)}]{Ren24}%
  \BibitemOpen
  \bibfield  {author} {\bibinfo {author} {\bibfnamefont {B.}~\bibnamefont
  {Hong}}\ and\ \bibinfo {author} {\bibfnamefont {Z.}~\bibnamefont {Ren}},\
  }\href {\doibase 10.1103/PhysRevD.109.023002} {\bibfield  {journal} {\bibinfo
   {journal} {Phys. Rev. D}\ }\textbf {\bibinfo {volume} {109}},\ \bibinfo
  {pages} {023002} (\bibinfo {year} {2024})}\BibitemShut {NoStop}%
\bibitem [{\citenamefont {Barbat}\ \emph {et~al.}(2024)\citenamefont {Barbat},
  \citenamefont {Schaffner-Bielich},\ and\ \citenamefont {Tolos}}]{Barbat24}%
  \BibitemOpen
  \bibfield  {author} {\bibinfo {author} {\bibfnamefont {M.~F.}\ \bibnamefont
  {Barbat}}, \bibinfo {author} {\bibfnamefont {J.}~\bibnamefont
  {Schaffner-Bielich}}, \ and\ \bibinfo {author} {\bibfnamefont
  {L.}~\bibnamefont {Tolos}},\ }\href {\doibase 10.1103/PhysRevD.110.023013}
  {\bibfield  {journal} {\bibinfo  {journal} {Phys. Rev. D}\ }\textbf {\bibinfo
  {volume} {110}},\ \bibinfo {pages} {023013} (\bibinfo {year}
  {2024})}\BibitemShut {NoStop}%
\bibitem [{\citenamefont {{Liu}}\ \emph {et~al.}(2026)\citenamefont {{Liu}},
  \citenamefont {{Chu}}, \citenamefont {{Liu}}, \citenamefont {{Li}},\ and\
  \citenamefont {{Li}}}]{Liu26}%
  \BibitemOpen
  \bibfield  {author} {\bibinfo {author} {\bibfnamefont {H.-M.}\ \bibnamefont
  {{Liu}}}, \bibinfo {author} {\bibfnamefont {P.-C.}\ \bibnamefont {{Chu}}},
  \bibinfo {author} {\bibfnamefont {H.}~\bibnamefont {{Liu}}}, \bibinfo
  {author} {\bibfnamefont {X.-H.}\ \bibnamefont {{Li}}}, \ and\ \bibinfo
  {author} {\bibfnamefont {Z.-H.}\ \bibnamefont {{Li}}},\ }\href {\doibase
  https://doi.org/10.1140/epjc/s10052-026-15893-9} {\bibfield  {journal}
  {\bibinfo  {journal} {Eur. Phys. J. C}\ }\textbf {\bibinfo {volume} {86}},\
  \bibinfo {pages} {649} (\bibinfo {year} {2026})}\BibitemShut {NoStop}%
\bibitem [{\citenamefont {{Bertone}}\ \emph {et~al.}(2005)\citenamefont
  {{Bertone}}, \citenamefont {{Hooper}},\ and\ \citenamefont
  {{Silk}}}]{Bertone05}%
  \BibitemOpen
  \bibfield  {author} {\bibinfo {author} {\bibfnamefont {G.}~\bibnamefont
  {{Bertone}}}, \bibinfo {author} {\bibfnamefont {D.}~\bibnamefont {{Hooper}}},
  \ and\ \bibinfo {author} {\bibfnamefont {J.}~\bibnamefont {{Silk}}},\ }\href
  {\doibase 10.1016/j.physrep.2004.08.031} {\bibfield  {journal} {\bibinfo
  {journal} {Phys. Rep.}\ }\textbf {\bibinfo {volume} {405}},\ \bibinfo {pages}
  {279} (\bibinfo {year} {2005})}\BibitemShut {NoStop}%
\bibitem [{\citenamefont {Massey}\ \emph {et~al.}(2010)\citenamefont {Massey},
  \citenamefont {Kitching},\ and\ \citenamefont {Richard}}]{Massey10}%
  \BibitemOpen
  \bibfield  {author} {\bibinfo {author} {\bibfnamefont {R.}~\bibnamefont
  {Massey}}, \bibinfo {author} {\bibfnamefont {T.}~\bibnamefont {Kitching}}, \
  and\ \bibinfo {author} {\bibfnamefont {J.}~\bibnamefont {Richard}},\ }\href
  {\doibase 10.1088/0034-4885/73/8/086901} {\bibfield  {journal} {\bibinfo
  {journal} {Rep. Prog. Phys.}\ }\textbf {\bibinfo {volume} {73}},\ \bibinfo
  {pages} {086901} (\bibinfo {year} {2010})}\BibitemShut {NoStop}%
\bibitem [{\citenamefont {Begeman}\ \emph {et~al.}(1991)\citenamefont
  {Begeman}, \citenamefont {Broeils},\ and\ \citenamefont
  {Sanders}}]{Begeman91}%
  \BibitemOpen
  \bibfield  {author} {\bibinfo {author} {\bibfnamefont {K.~G.}\ \bibnamefont
  {Begeman}}, \bibinfo {author} {\bibfnamefont {A.~H.}\ \bibnamefont
  {Broeils}}, \ and\ \bibinfo {author} {\bibfnamefont {R.~H.}\ \bibnamefont
  {Sanders}},\ }\href {\doibase 10.1093/mnras/249.3.523} {\bibfield  {journal}
  {\bibinfo  {journal} {Mon. Not. R. Astron. Soc.}\ }\textbf {\bibinfo {volume}
  {249}},\ \bibinfo {pages} {523} (\bibinfo {year} {1991})}\BibitemShut
  {NoStop}%
\bibitem [{\citenamefont {Wittman}\ \emph {et~al.}(2000)\citenamefont
  {Wittman}, \citenamefont {Tyson}, \citenamefont {Kirkman}, \citenamefont
  {Dell'Antonio},\ and\ \citenamefont {Bernstein}}]{Wittman00}%
  \BibitemOpen
  \bibfield  {author} {\bibinfo {author} {\bibfnamefont {D.~M.}\ \bibnamefont
  {Wittman}}, \bibinfo {author} {\bibfnamefont {J.~A.}\ \bibnamefont {Tyson}},
  \bibinfo {author} {\bibfnamefont {D.}~\bibnamefont {Kirkman}}, \bibinfo
  {author} {\bibfnamefont {I.}~\bibnamefont {Dell'Antonio}}, \ and\ \bibinfo
  {author} {\bibfnamefont {G.}~\bibnamefont {Bernstein}},\ }\href {\doibase
  10.1038/35012001} {\bibfield  {journal} {\bibinfo  {journal} {Nature
  (London)}\ }\textbf {\bibinfo {volume} {405}},\ \bibinfo {pages} {143}
  (\bibinfo {year} {2000})}\BibitemShut {NoStop}%
\bibitem [{\citenamefont {Abdalla}\ \emph {et~al.}(2010)\citenamefont
  {Abdalla}, \citenamefont {Abramo},\ and\ \citenamefont
  {de~Souza}}]{Abdalla10}%
  \BibitemOpen
  \bibfield  {author} {\bibinfo {author} {\bibfnamefont {E.}~\bibnamefont
  {Abdalla}}, \bibinfo {author} {\bibfnamefont {L.~R.}\ \bibnamefont {Abramo}},
  \ and\ \bibinfo {author} {\bibfnamefont {J.~C.~C.}\ \bibnamefont
  {de~Souza}},\ }\href {\doibase 10.1103/PhysRevD.82.023508} {\bibfield
  {journal} {\bibinfo  {journal} {Phys. Rev. D}\ }\textbf {\bibinfo {volume}
  {82}},\ \bibinfo {pages} {023508} (\bibinfo {year} {2010})}\BibitemShut
  {NoStop}%
\bibitem [{\citenamefont {Aprile}\ \emph {et~al.}(2022)\citenamefont {Aprile},
  \citenamefont {Abe}, \citenamefont {Agostini} \emph {et~al.}}]{Aprile22}%
  \BibitemOpen
  \bibfield  {author} {\bibinfo {author} {\bibfnamefont {E.}~\bibnamefont
  {Aprile}}, \bibinfo {author} {\bibfnamefont {K.}~\bibnamefont {Abe}},
  \bibinfo {author} {\bibfnamefont {F.}~\bibnamefont {Agostini}},  \emph
  {et~al.} (\bibinfo {collaboration} {XENON Collaboration}),\ }\href {\doibase
  10.1103/PhysRevLett.129.161805} {\bibfield  {journal} {\bibinfo  {journal}
  {Phys. Rev. Lett.}\ }\textbf {\bibinfo {volume} {129}},\ \bibinfo {pages}
  {161805} (\bibinfo {year} {2022})}\BibitemShut {NoStop}%
\bibitem [{\citenamefont {Agnes}\ \emph {et~al.}(2018)\citenamefont {Agnes}
  \emph {et~al.}}]{Agnes18}%
  \BibitemOpen
  \bibfield  {author} {\bibinfo {author} {\bibfnamefont {P.}~\bibnamefont
  {Agnes}} \emph {et~al.} (\bibinfo {collaboration} {The DarkSide
  Collaboration}),\ }\href {\doibase 10.1103/PhysRevLett.121.111303} {\bibfield
   {journal} {\bibinfo  {journal} {Phys. Rev. Lett.}\ }\textbf {\bibinfo
  {volume} {121}},\ \bibinfo {pages} {111303} (\bibinfo {year}
  {2018})}\BibitemShut {NoStop}%
\bibitem [{\citenamefont {Wang}\ \emph {et~al.}(2020)\citenamefont {Wang} \emph
  {et~al.}}]{Wang20}%
  \BibitemOpen
  \bibfield  {author} {\bibinfo {author} {\bibfnamefont {Q.}~\bibnamefont
  {Wang}} \emph {et~al.},\ }\href {\doibase 10.1088/1674-1137/abb658}
  {\bibfield  {journal} {\bibinfo  {journal} {Chin. Phys. C}\ }\textbf
  {\bibinfo {volume} {44}},\ \bibinfo {pages} {125001} (\bibinfo {year}
  {2020})}\BibitemShut {NoStop}%
\bibitem [{\citenamefont {Accardo}\ \emph {et~al.}(2014)\citenamefont {Accardo}
  \emph {et~al.}}]{Accardo14}%
  \BibitemOpen
  \bibfield  {author} {\bibinfo {author} {\bibfnamefont {L.}~\bibnamefont
  {Accardo}} \emph {et~al.} (\bibinfo {collaboration} {AMS Collaboration}),\
  }\href {\doibase 10.1103/PhysRevLett.113.121101} {\bibfield  {journal}
  {\bibinfo  {journal} {Phys. Rev. Lett.}\ }\textbf {\bibinfo {volume} {113}},\
  \bibinfo {pages} {121101} (\bibinfo {year} {2014})}\BibitemShut {NoStop}%
\bibitem [{\citenamefont {Atwood}\ \emph {et~al.}(2009)\citenamefont {Atwood}
  \emph {et~al.}}]{Atwood09}%
  \BibitemOpen
  \bibfield  {author} {\bibinfo {author} {\bibfnamefont {W.~B.}\ \bibnamefont
  {Atwood}} \emph {et~al.},\ }\href {\doibase 10.1088/0004-637X/697/2/1071}
  {\bibfield  {journal} {\bibinfo  {journal} {\apj}\ }\textbf {\bibinfo
  {volume} {697}},\ \bibinfo {pages} {1071} (\bibinfo {year}
  {2009})}\BibitemShut {NoStop}%
\bibitem [{\citenamefont {Adriani}\ \emph {et~al.}(2018)\citenamefont {Adriani}
  \emph {et~al.}}]{Adriani18}%
  \BibitemOpen
  \bibfield  {author} {\bibinfo {author} {\bibfnamefont {O.}~\bibnamefont
  {Adriani}} \emph {et~al.} (\bibinfo {collaboration} {CALET Collaboration}),\
  }\href {\doibase 10.1103/PhysRevLett.120.261102} {\bibfield  {journal}
  {\bibinfo  {journal} {Phys. Rev. Lett.}\ }\textbf {\bibinfo {volume} {120}},\
  \bibinfo {pages} {261102} (\bibinfo {year} {2018})}\BibitemShut {NoStop}%
\bibitem [{\citenamefont {Sato}\ and\ \citenamefont {Tobioka}(2016)}]{Sato16}%
  \BibitemOpen
  \bibfield  {author} {\bibinfo {author} {\bibfnamefont {R.}~\bibnamefont
  {Sato}}\ and\ \bibinfo {author} {\bibfnamefont {K.}~\bibnamefont {Tobioka}},\
  }\href {\doibase 10.1016/j.physletb.2016.07.051} {\bibfield  {journal}
  {\bibinfo  {journal} {Phys. Lett. B}\ }\textbf {\bibinfo {volume} {760}},\
  \bibinfo {pages} {590} (\bibinfo {year} {2016})}\BibitemShut {NoStop}%
\bibitem [{\citenamefont {{Shakeri}}\ and\ \citenamefont
  {{Karkevandi}}(2024)}]{Shakeri24}%
  \BibitemOpen
  \bibfield  {author} {\bibinfo {author} {\bibfnamefont {S.}~\bibnamefont
  {{Shakeri}}}\ and\ \bibinfo {author} {\bibfnamefont {D.~R.}\ \bibnamefont
  {{Karkevandi}}},\ }\href {\doibase 10.1103/PhysRevD.109.043029} {\bibfield
  {journal} {\bibinfo  {journal} {\prd}\ }\textbf {\bibinfo {volume} {109}},\
  \bibinfo {eid} {043029} (\bibinfo {year} {2024})}\BibitemShut {NoStop}%
\bibitem [{\citenamefont {Kumar}\ and\ \citenamefont {Sotani}(2024)}]{Kumar24}%
  \BibitemOpen
  \bibfield  {author} {\bibinfo {author} {\bibfnamefont {A.}~\bibnamefont
  {Kumar}}\ and\ \bibinfo {author} {\bibfnamefont {H.}~\bibnamefont {Sotani}},\
  }\href {\doibase 10.1103/PhysRevD.110.063001} {\bibfield  {journal} {\bibinfo
   {journal} {Phys. Rev. D}\ }\textbf {\bibinfo {volume} {110}},\ \bibinfo
  {pages} {063001} (\bibinfo {year} {2024})}\BibitemShut {NoStop}%
\bibitem [{\citenamefont {{Bell}}\ \emph {et~al.}(2021)\citenamefont {{Bell}},
  \citenamefont {{Busoni}}, \citenamefont {{Motta}}, \citenamefont {{Robles}},
  \citenamefont {{Thomas}},\ and\ \citenamefont {{Virgato}}}]{Bell21}%
  \BibitemOpen
  \bibfield  {author} {\bibinfo {author} {\bibfnamefont {N.~F.}\ \bibnamefont
  {{Bell}}}, \bibinfo {author} {\bibfnamefont {G.}~\bibnamefont {{Busoni}}},
  \bibinfo {author} {\bibfnamefont {T.~F.}\ \bibnamefont {{Motta}}}, \bibinfo
  {author} {\bibfnamefont {S.}~\bibnamefont {{Robles}}}, \bibinfo {author}
  {\bibfnamefont {A.~W.}\ \bibnamefont {{Thomas}}}, \ and\ \bibinfo {author}
  {\bibfnamefont {M.}~\bibnamefont {{Virgato}}},\ }\href {\doibase
  10.1103/PhysRevLett.127.111803} {\bibfield  {journal} {\bibinfo  {journal}
  {Phys. Rev. Lett.}\ }\textbf {\bibinfo {volume} {127}},\ \bibinfo {eid}
  {111803} (\bibinfo {year} {2021})}\BibitemShut {NoStop}%
\bibitem [{\citenamefont {{Collier}}\ \emph {et~al.}(2022)\citenamefont
  {{Collier}}, \citenamefont {{Croon}},\ and\ \citenamefont
  {{Leane}}}]{Collier22}%
  \BibitemOpen
  \bibfield  {author} {\bibinfo {author} {\bibfnamefont {M.}~\bibnamefont
  {{Collier}}}, \bibinfo {author} {\bibfnamefont {D.}~\bibnamefont {{Croon}}},
  \ and\ \bibinfo {author} {\bibfnamefont {R.~K.}\ \bibnamefont {{Leane}}},\
  }\href {\doibase 10.1103/PhysRevD.106.123027} {\bibfield  {journal} {\bibinfo
   {journal} {\prd}\ }\textbf {\bibinfo {volume} {106}},\ \bibinfo {eid}
  {123027} (\bibinfo {year} {2022})}\BibitemShut {NoStop}%
\bibitem [{\citenamefont {{Gleason}}\ \emph {et~al.}(2022)\citenamefont
  {{Gleason}}, \citenamefont {{Brown}},\ and\ \citenamefont
  {{Kain}}}]{Gleason22}%
  \BibitemOpen
  \bibfield  {author} {\bibinfo {author} {\bibfnamefont {T.}~\bibnamefont
  {{Gleason}}}, \bibinfo {author} {\bibfnamefont {B.}~\bibnamefont {{Brown}}},
  \ and\ \bibinfo {author} {\bibfnamefont {B.}~\bibnamefont {{Kain}}},\ }\href
  {\doibase 10.1103/PhysRevD.105.023010} {\bibfield  {journal} {\bibinfo
  {journal} {Phys. Rev. D}\ }\textbf {\bibinfo {volume} {105}},\ \bibinfo {eid}
  {023010} (\bibinfo {year} {2022})}\BibitemShut {NoStop}%
\bibitem [{\citenamefont {{Xiang}}\ \emph {et~al.}(2014)\citenamefont
  {{Xiang}}, \citenamefont {{Jiang}}, \citenamefont {{Zhang}},\ and\
  \citenamefont {{Yang}}}]{Xiang14}%
  \BibitemOpen
  \bibfield  {author} {\bibinfo {author} {\bibfnamefont {Q.-F.}\ \bibnamefont
  {{Xiang}}}, \bibinfo {author} {\bibfnamefont {W.-Z.}\ \bibnamefont
  {{Jiang}}}, \bibinfo {author} {\bibfnamefont {D.-R.}\ \bibnamefont
  {{Zhang}}}, \ and\ \bibinfo {author} {\bibfnamefont {R.-Y.}\ \bibnamefont
  {{Yang}}},\ }\href {\doibase 10.1103/PhysRevC.89.025803} {\bibfield
  {journal} {\bibinfo  {journal} {Phys. Rev. C}\ }\textbf {\bibinfo {volume}
  {89}},\ \bibinfo {eid} {025803} (\bibinfo {year} {2014})}\BibitemShut
  {NoStop}%
\bibitem [{\citenamefont {Tolos}\ and\ \citenamefont
  {Schaffner-Bielich}(2015)}]{Tolos15}%
  \BibitemOpen
  \bibfield  {author} {\bibinfo {author} {\bibfnamefont {L.}~\bibnamefont
  {Tolos}}\ and\ \bibinfo {author} {\bibfnamefont {J.}~\bibnamefont
  {Schaffner-Bielich}},\ }\href {\doibase 10.1103/PhysRevD.92.123002}
  {\bibfield  {journal} {\bibinfo  {journal} {Phys. Rev. D}\ }\textbf {\bibinfo
  {volume} {92}},\ \bibinfo {pages} {123002} (\bibinfo {year}
  {2015})}\BibitemShut {NoStop}%
\bibitem [{\citenamefont {Das}\ \emph {et~al.}(2019)\citenamefont {Das},
  \citenamefont {Malik},\ and\ \citenamefont {Nayak}}]{ADas19}%
  \BibitemOpen
  \bibfield  {author} {\bibinfo {author} {\bibfnamefont {A.}~\bibnamefont
  {Das}}, \bibinfo {author} {\bibfnamefont {T.}~\bibnamefont {Malik}}, \ and\
  \bibinfo {author} {\bibfnamefont {A.~C.}\ \bibnamefont {Nayak}},\ }\href
  {\doibase 10.1103/PhysRevD.99.043016} {\bibfield  {journal} {\bibinfo
  {journal} {Phys. Rev. D}\ }\textbf {\bibinfo {volume} {99}},\ \bibinfo
  {pages} {043016} (\bibinfo {year} {2019})}\BibitemShut {NoStop}%
\bibitem [{\citenamefont {Yang}\ \emph {et~al.}(2021)\citenamefont {Yang},
  \citenamefont {Pi},\ and\ \citenamefont {Zheng}}]{Yang21}%
  \BibitemOpen
  \bibfield  {author} {\bibinfo {author} {\bibfnamefont {S.-H.}\ \bibnamefont
  {Yang}}, \bibinfo {author} {\bibfnamefont {C.-M.}\ \bibnamefont {Pi}}, \ and\
  \bibinfo {author} {\bibfnamefont {X.-P.}\ \bibnamefont {Zheng}},\ }\href
  {\doibase 10.1103/PhysRevD.104.083016} {\bibfield  {journal} {\bibinfo
  {journal} {Phys. Rev. D}\ }\textbf {\bibinfo {volume} {104}},\ \bibinfo
  {pages} {083016} (\bibinfo {year} {2021})}\BibitemShut {NoStop}%
\bibitem [{\citenamefont {Kain}(2021)}]{Kain21}%
  \BibitemOpen
  \bibfield  {author} {\bibinfo {author} {\bibfnamefont {B.}~\bibnamefont
  {Kain}},\ }\href {\doibase 10.1103/PhysRevD.103.043009} {\bibfield  {journal}
  {\bibinfo  {journal} {Phys. Rev. D}\ }\textbf {\bibinfo {volume} {103}},\
  \bibinfo {pages} {043009} (\bibinfo {year} {2021})}\BibitemShut {NoStop}%
\bibitem [{\citenamefont {{Emma}}\ \emph {et~al.}(2022)\citenamefont {{Emma}},
  \citenamefont {{Schianchi}}, \citenamefont {{Pannarale}}, \citenamefont
  {{Sagun}},\ and\ \citenamefont {{Dietrich}}}]{Emma22}%
  \BibitemOpen
  \bibfield  {author} {\bibinfo {author} {\bibfnamefont {M.}~\bibnamefont
  {{Emma}}}, \bibinfo {author} {\bibfnamefont {F.}~\bibnamefont {{Schianchi}}},
  \bibinfo {author} {\bibfnamefont {F.}~\bibnamefont {{Pannarale}}}, \bibinfo
  {author} {\bibfnamefont {V.}~\bibnamefont {{Sagun}}}, \ and\ \bibinfo
  {author} {\bibfnamefont {T.}~\bibnamefont {{Dietrich}}},\ }\href {\doibase
  10.3390/particles5030024} {\bibfield  {journal} {\bibinfo  {journal}
  {Particles}\ }\textbf {\bibinfo {volume} {5}},\ \bibinfo {pages} {273}
  (\bibinfo {year} {2022})}\BibitemShut {NoStop}%
\bibitem [{\citenamefont {{Bauswein}}\ \emph {et~al.}(2023)\citenamefont
  {{Bauswein}}, \citenamefont {{Guo}}, \citenamefont {{Lien-Hua}},
  \citenamefont {{Lin}},\ and\ \citenamefont {{Wu}}}]{Bauswein23}%
  \BibitemOpen
  \bibfield  {author} {\bibinfo {author} {\bibfnamefont {A.}~\bibnamefont
  {{Bauswein}}}, \bibinfo {author} {\bibfnamefont {G.}~\bibnamefont {{Guo}}},
  \bibinfo {author} {\bibfnamefont {J.}~\bibnamefont {{Lien-Hua}}}, \bibinfo
  {author} {\bibfnamefont {Y.-H.}\ \bibnamefont {{Lin}}}, \ and\ \bibinfo
  {author} {\bibfnamefont {M.-R.}\ \bibnamefont {{Wu}}},\ }\href {\doibase
  10.1103/PhysRevD.107.083002} {\bibfield  {journal} {\bibinfo  {journal}
  {\prd}\ }\textbf {\bibinfo {volume} {107}},\ \bibinfo {eid} {083002}
  (\bibinfo {year} {2023})}\BibitemShut {NoStop}%
\bibitem [{\citenamefont {Routaray}\ \emph
  {et~al.}(2023{\natexlab{b}})\citenamefont {Routaray}, \citenamefont {Quddus},
  \citenamefont {Chakravarti},\ and\ \citenamefont {Kumar}}]{Routaray23b}%
  \BibitemOpen
  \bibfield  {author} {\bibinfo {author} {\bibfnamefont {P.}~\bibnamefont
  {Routaray}}, \bibinfo {author} {\bibfnamefont {A.}~\bibnamefont {Quddus}},
  \bibinfo {author} {\bibfnamefont {K.}~\bibnamefont {Chakravarti}}, \ and\
  \bibinfo {author} {\bibfnamefont {B.}~\bibnamefont {Kumar}},\ }\href
  {\doibase 10.1093/mnras/stad2628} {\bibfield  {journal} {\bibinfo  {journal}
  {Mon. Not. R. Astron. Soc.}\ }\textbf {\bibinfo {volume} {525}},\ \bibinfo
  {pages} {5492} (\bibinfo {year} {2023}{\natexlab{b}})}\BibitemShut {NoStop}%
\bibitem [{\citenamefont {{Liu}}\ \emph {et~al.}(2023)\citenamefont {{Liu}},
  \citenamefont {{Wei}}, \citenamefont {{Li}}, \citenamefont {{Burgio}},\ and\
  \citenamefont {{Schulze}}}]{Liu23}%
  \BibitemOpen
  \bibfield  {author} {\bibinfo {author} {\bibfnamefont {H.-M.}\ \bibnamefont
  {{Liu}}}, \bibinfo {author} {\bibfnamefont {J.-B.}\ \bibnamefont {{Wei}}},
  \bibinfo {author} {\bibfnamefont {Z.-H.}\ \bibnamefont {{Li}}}, \bibinfo
  {author} {\bibfnamefont {G.~F.}\ \bibnamefont {{Burgio}}}, \ and\ \bibinfo
  {author} {\bibfnamefont {H.~J.}\ \bibnamefont {{Schulze}}},\ }\href {\doibase
  10.1016/j.dark.2023.101338} {\bibfield  {journal} {\bibinfo  {journal} {Phys.
  Dark Universe}\ }\textbf {\bibinfo {volume} {42}},\ \bibinfo {eid} {101338}
  (\bibinfo {year} {2023})}\BibitemShut {NoStop}%
\bibitem [{\citenamefont {Yang}\ and\ \citenamefont {Yang}(2025)}]{Wu25}%
  \BibitemOpen
  \bibfield  {author} {\bibinfo {author} {\bibfnamefont {W.~C.}\ \bibnamefont
  {Yang}, \bibfnamefont {Yida}}\ and\ \bibinfo {author} {\bibfnamefont {J.-F.}\
  \bibnamefont {Yang}},\ }\href {\doibase 10.1140/epjc/s10052-025-14150-9}
  {\bibfield  {journal} {\bibinfo  {journal} {Eur. Phys. J. C}\ }\textbf
  {\bibinfo {volume} {85}},\ \bibinfo {pages} {426} (\bibinfo {year}
  {2025})}\BibitemShut {NoStop}%
\bibitem [{\citenamefont {Liu}\ \emph {et~al.}(2024)\citenamefont {Liu},
  \citenamefont {Wei}, \citenamefont {Li}, \citenamefont {Burgio},
  \citenamefont {Das},\ and\ \citenamefont {Schulze}}]{Liu24}%
  \BibitemOpen
  \bibfield  {author} {\bibinfo {author} {\bibfnamefont {H.-M.}\ \bibnamefont
  {Liu}}, \bibinfo {author} {\bibfnamefont {J.-B.}\ \bibnamefont {Wei}},
  \bibinfo {author} {\bibfnamefont {Z.-H.}\ \bibnamefont {Li}}, \bibinfo
  {author} {\bibfnamefont {G.~F.}\ \bibnamefont {Burgio}}, \bibinfo {author}
  {\bibfnamefont {H.~C.}\ \bibnamefont {Das}}, \ and\ \bibinfo {author}
  {\bibfnamefont {H.-J.}\ \bibnamefont {Schulze}},\ }\href {\doibase
  10.1103/PhysRevD.110.023024} {\bibfield  {journal} {\bibinfo  {journal}
  {Phys. Rev. D}\ }\textbf {\bibinfo {volume} {110}},\ \bibinfo {pages}
  {023024} (\bibinfo {year} {2024})}\BibitemShut {NoStop}%
\bibitem [{\citenamefont {Sedrakian}(2019)}]{Sedrakian19}%
  \BibitemOpen
  \bibfield  {author} {\bibinfo {author} {\bibfnamefont {A.}~\bibnamefont
  {Sedrakian}},\ }\href {\doibase 10.1103/PhysRevD.99.043011} {\bibfield
  {journal} {\bibinfo  {journal} {Phys. Rev. D}\ }\textbf {\bibinfo {volume}
  {99}},\ \bibinfo {pages} {043011} (\bibinfo {year} {2019})}\BibitemShut
  {NoStop}%
\bibitem [{\citenamefont {{Zhang}}\ \emph {et~al.}(2022)\citenamefont
  {{Zhang}}, \citenamefont {{Huang}}, \citenamefont {{Tsao}},\ and\
  \citenamefont {{Lin}}}]{Zhang22}%
  \BibitemOpen
  \bibfield  {author} {\bibinfo {author} {\bibfnamefont {K.}~\bibnamefont
  {{Zhang}}}, \bibinfo {author} {\bibfnamefont {G.-Z.}\ \bibnamefont
  {{Huang}}}, \bibinfo {author} {\bibfnamefont {J.-S.}\ \bibnamefont {{Tsao}}},
  \ and\ \bibinfo {author} {\bibfnamefont {F.-L.}\ \bibnamefont {{Lin}}},\
  }\href {\doibase 10.1140/epjc/s10052-022-10335-8} {\bibfield  {journal}
  {\bibinfo  {journal} {EPJC}\ }\textbf {\bibinfo {volume} {82}},\ \bibinfo
  {eid} {366} (\bibinfo {year} {2022})}\BibitemShut {NoStop}%
\bibitem [{\citenamefont {Singh}\ \emph {et~al.}(2023)\citenamefont {Singh},
  \citenamefont {Gupta}, \citenamefont {Berti}, \citenamefont {Reddy},\ and\
  \citenamefont {Sathyaprakash}}]{Gupta23}%
  \BibitemOpen
  \bibfield  {author} {\bibinfo {author} {\bibfnamefont {D.}~\bibnamefont
  {Singh}}, \bibinfo {author} {\bibfnamefont {A.}~\bibnamefont {Gupta}},
  \bibinfo {author} {\bibfnamefont {E.}~\bibnamefont {Berti}}, \bibinfo
  {author} {\bibfnamefont {S.}~\bibnamefont {Reddy}}, \ and\ \bibinfo {author}
  {\bibfnamefont {B.~S.}\ \bibnamefont {Sathyaprakash}},\ }\href {\doibase
  10.1103/PhysRevD.107.083037} {\bibfield  {journal} {\bibinfo  {journal}
  {Phys. Rev. D}\ }\textbf {\bibinfo {volume} {107}},\ \bibinfo {pages}
  {083037} (\bibinfo {year} {2023})}\BibitemShut {NoStop}%
\bibitem [{\citenamefont {Kini}\ \emph {et~al.}(2024)\citenamefont {Kini} \emph
  {et~al.}}]{Kini24}%
  \BibitemOpen
  \bibfield  {author} {\bibinfo {author} {\bibfnamefont {Y.}~\bibnamefont
  {Kini}} \emph {et~al.},\ }\href {\doibase 10.1093/mnras/stae2398} {\bibfield
  {journal} {\bibinfo  {journal} {\mnras}\ }\textbf {\bibinfo {volume} {535}},\
  \bibinfo {pages} {1507} (\bibinfo {year} {2024})}\BibitemShut {NoStop}%
\bibitem [{\citenamefont {Pitz}\ and\ \citenamefont
  {Schaffner-Bielich}(2025)}]{Pitz25}%
  \BibitemOpen
  \bibfield  {author} {\bibinfo {author} {\bibfnamefont {S.~L.}\ \bibnamefont
  {Pitz}}\ and\ \bibinfo {author} {\bibfnamefont {J.}~\bibnamefont
  {Schaffner-Bielich}},\ }\href {\doibase 10.1103/PhysRevD.111.043050}
  {\bibfield  {journal} {\bibinfo  {journal} {Phys. Rev. D}\ }\textbf {\bibinfo
  {volume} {111}},\ \bibinfo {pages} {043050} (\bibinfo {year}
  {2025})}\BibitemShut {NoStop}%
\bibitem [{\citenamefont {Yang}\ \emph {et~al.}(2025)\citenamefont {Yang},
  \citenamefont {Pi},\ and\ \citenamefont {Weber}}]{Yang25}%
  \BibitemOpen
  \bibfield  {author} {\bibinfo {author} {\bibfnamefont {S.-H.}\ \bibnamefont
  {Yang}}, \bibinfo {author} {\bibfnamefont {C.-M.}\ \bibnamefont {Pi}}, \ and\
  \bibinfo {author} {\bibfnamefont {F.}~\bibnamefont {Weber}},\ }\href
  {\doibase 10.1103/PhysRevD.111.043037} {\bibfield  {journal} {\bibinfo
  {journal} {Phys. Rev. D}\ }\textbf {\bibinfo {volume} {111}},\ \bibinfo
  {pages} {043037} (\bibinfo {year} {2025})}\BibitemShut {NoStop}%
\bibitem [{\citenamefont {Lopes}\ and\ \citenamefont {Issifu}(2025)}]{Lopes25}%
  \BibitemOpen
  \bibfield  {author} {\bibinfo {author} {\bibfnamefont {L.~L.}\ \bibnamefont
  {Lopes}}\ and\ \bibinfo {author} {\bibfnamefont {A.}~\bibnamefont {Issifu}},\
  }\href {\doibase https://doi.org/10.1016/j.dark.2025.101922} {\bibfield
  {journal} {\bibinfo  {journal} {Phys. Dark Universe}\ }\textbf {\bibinfo
  {volume} {48}},\ \bibinfo {pages} {101922} (\bibinfo {year}
  {2025})}\BibitemShut {NoStop}%
\bibitem [{\citenamefont {Wiringa}\ \emph {et~al.}(1995)\citenamefont
  {Wiringa}, \citenamefont {Stoks},\ and\ \citenamefont
  {Schiavilla}}]{Wiringa95}%
  \BibitemOpen
  \bibfield  {author} {\bibinfo {author} {\bibfnamefont {R.~B.}\ \bibnamefont
  {Wiringa}}, \bibinfo {author} {\bibfnamefont {V.}~\bibnamefont {Stoks}}, \
  and\ \bibinfo {author} {\bibfnamefont {R.}~\bibnamefont {Schiavilla}},\
  }\href {\doibase 10.1103/PhysRevC.51.38} {\bibfield  {journal} {\bibinfo
  {journal} {Phys. Rev. C}\ }\textbf {\bibinfo {volume} {51}},\ \bibinfo
  {pages} {38} (\bibinfo {year} {1995})}\BibitemShut {NoStop}%
%%CITATION = NUCL-TH/9408016;%%
\bibitem [{\citenamefont {Grang\'e}\ \emph {et~al.}(1989)\citenamefont
  {Grang\'e}, \citenamefont {Lejeune}, \citenamefont {Martzolff},\ and\
  \citenamefont {Mathiot}}]{Grange89}%
  \BibitemOpen
  \bibfield  {author} {\bibinfo {author} {\bibfnamefont {P.}~\bibnamefont
  {Grang\'e}}, \bibinfo {author} {\bibfnamefont {A.}~\bibnamefont {Lejeune}},
  \bibinfo {author} {\bibfnamefont {M.}~\bibnamefont {Martzolff}}, \ and\
  \bibinfo {author} {\bibfnamefont {J.-F.}\ \bibnamefont {Mathiot}},\ }\href
  {\doibase 10.1103/PhysRevC.40.1040} {\bibfield  {journal} {\bibinfo
  {journal} {Phys. Rev. C}\ }\textbf {\bibinfo {volume} {40}},\ \bibinfo
  {pages} {1040} (\bibinfo {year} {1989})}\BibitemShut {NoStop}%
\bibitem [{\citenamefont {{Zuo}}\ \emph {et~al.}(2002)\citenamefont {{Zuo}},
  \citenamefont {{Lejeune}}, \citenamefont {{Lombardo}},\ and\ \citenamefont
  {{Mathiot}}}]{Zuo02}%
  \BibitemOpen
  \bibfield  {author} {\bibinfo {author} {\bibfnamefont {W.}~\bibnamefont
  {{Zuo}}}, \bibinfo {author} {\bibfnamefont {A.}~\bibnamefont {{Lejeune}}},
  \bibinfo {author} {\bibfnamefont {U.}~\bibnamefont {{Lombardo}}}, \ and\
  \bibinfo {author} {\bibfnamefont {J.~F.}\ \bibnamefont {{Mathiot}}},\ }\href
  {\doibase 10.1140/epja/i2002-10031-y} {\bibfield  {journal} {\bibinfo
  {journal} {Eur. Phys. J. A}\ }\textbf {\bibinfo {volume} {14}},\ \bibinfo
  {pages} {469} (\bibinfo {year} {2002})}\BibitemShut {NoStop}%
\bibitem [{\citenamefont {{Li}}\ \emph {et~al.}(2008)\citenamefont {{Li}},
  \citenamefont {{Lombardo}}, \citenamefont {{Schulze}},\ and\ \citenamefont
  {{Zuo}}}]{Li08a}%
  \BibitemOpen
  \bibfield  {author} {\bibinfo {author} {\bibfnamefont {Z.~H.}\ \bibnamefont
  {{Li}}}, \bibinfo {author} {\bibfnamefont {U.}~\bibnamefont {{Lombardo}}},
  \bibinfo {author} {\bibfnamefont {H.-J.}\ \bibnamefont {{Schulze}}}, \ and\
  \bibinfo {author} {\bibfnamefont {W.}~\bibnamefont {{Zuo}}},\ }\href
  {\doibase 10.1103/PhysRevC.77.034316} {\bibfield  {journal} {\bibinfo
  {journal} {Phys. Rev. C}\ }\textbf {\bibinfo {volume} {77}},\ \bibinfo {eid}
  {034316} (\bibinfo {year} {2008})}\BibitemShut {NoStop}%
\bibitem [{\citenamefont {{Li}}\ and\ \citenamefont {{Schulze}}(2008)}]{Li08b}%
  \BibitemOpen
  \bibfield  {author} {\bibinfo {author} {\bibfnamefont {Z.~H.}\ \bibnamefont
  {{Li}}}\ and\ \bibinfo {author} {\bibfnamefont {H.-J.}\ \bibnamefont
  {{Schulze}}},\ }\href {\doibase 10.1103/PhysRevC.78.028801} {\bibfield
  {journal} {\bibinfo  {journal} {Phys. Rev. C}\ }\textbf {\bibinfo {volume}
  {78}},\ \bibinfo {eid} {028801} (\bibinfo {year} {2008})}\BibitemShut
  {NoStop}%
\bibitem [{\citenamefont {{Wei}}\ \emph {et~al.}(2020)\citenamefont {{Wei}},
  \citenamefont {{Lu}}, \citenamefont {{Burgio}}, \citenamefont {{Li}},\ and\
  \citenamefont {{Schulze}}}]{Wei20}%
  \BibitemOpen
  \bibfield  {author} {\bibinfo {author} {\bibfnamefont {J.-B.}\ \bibnamefont
  {{Wei}}}, \bibinfo {author} {\bibfnamefont {J.-J.}\ \bibnamefont {{Lu}}},
  \bibinfo {author} {\bibfnamefont {G.~F.}\ \bibnamefont {{Burgio}}}, \bibinfo
  {author} {\bibfnamefont {Z.-H.}\ \bibnamefont {{Li}}}, \ and\ \bibinfo
  {author} {\bibfnamefont {H.-J.}\ \bibnamefont {{Schulze}}},\ }\href {\doibase
  10.1140/epja/s10050-020-00058-3} {\bibfield  {journal} {\bibinfo  {journal}
  {Eur. Phys. J. A}\ }\textbf {\bibinfo {volume} {56}},\ \bibinfo {eid} {63}
  (\bibinfo {year} {2020})}\BibitemShut {NoStop}%
\bibitem [{\citenamefont {{Burgio}}\ \emph {et~al.}(2021)\citenamefont
  {{Burgio}}, \citenamefont {{Schulze}}, \citenamefont {{Vida{\~n}a}},\ and\
  \citenamefont {{Wei}}}]{Burgio21b}%
  \BibitemOpen
  \bibfield  {author} {\bibinfo {author} {\bibfnamefont {G.~F.}\ \bibnamefont
  {{Burgio}}}, \bibinfo {author} {\bibfnamefont {H.-J.}\ \bibnamefont
  {{Schulze}}}, \bibinfo {author} {\bibfnamefont {I.}~\bibnamefont
  {{Vida{\~n}a}}}, \ and\ \bibinfo {author} {\bibfnamefont {J.-B.}\
  \bibnamefont {{Wei}}},\ }\href {\doibase 10.3390/sym13030400} {\bibfield
  {journal} {\bibinfo  {journal} {Symmetry}\ }\textbf {\bibinfo {volume}
  {13}},\ \bibinfo {pages} {400} (\bibinfo {year} {2021})}\BibitemShut
  {NoStop}%
\bibitem [{\citenamefont {Shen}\ \emph {et~al.}(2020)\citenamefont {Shen},
  \citenamefont {Ji}, \citenamefont {Hu},\ and\ \citenamefont
  {Sumiyoshi}}]{Shen20}%
  \BibitemOpen
  \bibfield  {author} {\bibinfo {author} {\bibfnamefont {H.}~\bibnamefont
  {Shen}}, \bibinfo {author} {\bibfnamefont {F.}~\bibnamefont {Ji}}, \bibinfo
  {author} {\bibfnamefont {J.}~\bibnamefont {Hu}}, \ and\ \bibinfo {author}
  {\bibfnamefont {K.}~\bibnamefont {Sumiyoshi}},\ }\href {\doibase
  10.3847/1538-4357/ab72fd} {\bibfield  {journal} {\bibinfo  {journal} {\apj}\
  }\textbf {\bibinfo {volume} {891}},\ \bibinfo {pages} {148} (\bibinfo {year}
  {2020})}\BibitemShut {NoStop}%
\bibitem [{\citenamefont {Tolman}(1939)}]{Tolman39}%
  \BibitemOpen
  \bibfield  {author} {\bibinfo {author} {\bibfnamefont {R.~C.}\ \bibnamefont
  {Tolman}},\ }\href {\doibase 10.1103/PhysRev.55.364} {\bibfield  {journal}
  {\bibinfo  {journal} {Phys. Rev.}\ }\textbf {\bibinfo {volume} {55}},\
  \bibinfo {pages} {364} (\bibinfo {year} {1939})}\BibitemShut {NoStop}%
\bibitem [{\citenamefont {Oppenheimer}\ and\ \citenamefont
  {Volkoff}(1939)}]{Oppenheimer39}%
  \BibitemOpen
  \bibfield  {author} {\bibinfo {author} {\bibfnamefont {J.~R.}\ \bibnamefont
  {Oppenheimer}}\ and\ \bibinfo {author} {\bibfnamefont {G.~M.}\ \bibnamefont
  {Volkoff}},\ }\href {\doibase 10.1103/PhysRev.55.374} {\bibfield  {journal}
  {\bibinfo  {journal} {Phys. Rev.}\ }\textbf {\bibinfo {volume} {55}},\
  \bibinfo {pages} {374} (\bibinfo {year} {1939})}\BibitemShut {NoStop}%
\bibitem [{\citenamefont {{Romani}}\ \emph {et~al.}(2026)\citenamefont
  {{Romani}}, \citenamefont {{Beleznay}}, \citenamefont {{Filippenko}},
  \citenamefont {{Brink}},\ and\ \citenamefont {{Zheng}}}]{Romani25}%
  \BibitemOpen
  \bibfield  {author} {\bibinfo {author} {\bibfnamefont {R.~W.}\ \bibnamefont
  {{Romani}}}, \bibinfo {author} {\bibfnamefont {M.}~\bibnamefont
  {{Beleznay}}}, \bibinfo {author} {\bibfnamefont {A.~V.}\ \bibnamefont
  {{Filippenko}}}, \bibinfo {author} {\bibfnamefont {T.~G.}\ \bibnamefont
  {{Brink}}}, \ and\ \bibinfo {author} {\bibfnamefont {W.}~\bibnamefont
  {{Zheng}}},\ }\href {\doibase 10.3847/1538-4357/ae28c5} {\bibfield  {journal}
  {\bibinfo  {journal} {\apj}\ }\textbf {\bibinfo {volume} {996}},\ \bibinfo
  {eid} {101} (\bibinfo {year} {2026})}\BibitemShut {NoStop}%
\bibitem [{\citenamefont {{Vinciguerra}}\ \emph {et~al.}(2024)\citenamefont
  {{Vinciguerra}} \emph {et~al.}}]{Vinciguerra24}%
  \BibitemOpen
  \bibfield  {author} {\bibinfo {author} {\bibfnamefont {S.}~\bibnamefont
  {{Vinciguerra}}} \emph {et~al.},\ }\href {\doibase 10.3847/1538-4357/acfb83}
  {\bibfield  {journal} {\bibinfo  {journal} {Astrophys. J.}\ }\textbf
  {\bibinfo {volume} {961}},\ \bibinfo {eid} {62} (\bibinfo {year}
  {2024})}\BibitemShut {NoStop}%
\bibitem [{\citenamefont {{Salmi}}\ \emph {et~al.}(2024)\citenamefont {{Salmi}}
  \emph {et~al.}}]{Salmi24}%
  \BibitemOpen
  \bibfield  {author} {\bibinfo {author} {\bibfnamefont {T.}~\bibnamefont
  {{Salmi}}} \emph {et~al.},\ }\href {\doibase 10.3847/1538-4357/ad5f1f}
  {\bibfield  {journal} {\bibinfo  {journal} {Astrophys. J.}\ }\textbf
  {\bibinfo {volume} {974}},\ \bibinfo {eid} {294} (\bibinfo {year}
  {2024})}\BibitemShut {NoStop}%
\bibitem [{\citenamefont {{Miller}}\ \emph {et~al.}(2026)\citenamefont
  {{Miller}} \emph {et~al.}}]{Miller26}%
  \BibitemOpen
  \bibfield  {author} {\bibinfo {author} {\bibfnamefont {M.~C.}\ \bibnamefont
  {{Miller}}} \emph {et~al.},\ }\href {\doibase 10.3847/2041-8213/ae5057}
  {\bibfield  {journal} {\bibinfo  {journal} {\apjl}\ }\textbf {\bibinfo
  {volume} {1000}},\ \bibinfo {eid} {L48} (\bibinfo {year} {2026})}\BibitemShut
  {NoStop}%
\bibitem [{\citenamefont {{Raaijmakers}}\ \emph {et~al.}(2021)\citenamefont
  {{Raaijmakers}} \emph {et~al.}}]{Raaijmakers21}%
  \BibitemOpen
  \bibfield  {author} {\bibinfo {author} {\bibfnamefont {G.}~\bibnamefont
  {{Raaijmakers}}} \emph {et~al.},\ }\href {\doibase 10.3847/2041-8213/ac089a}
  {\bibfield  {journal} {\bibinfo  {journal} {Astrophys. J. Lett.}\ }\textbf
  {\bibinfo {volume} {918}},\ \bibinfo {eid} {L29} (\bibinfo {year}
  {2021})}\BibitemShut {NoStop}%
\bibitem [{\citenamefont {{Pang}}\ \emph {et~al.}(2021)\citenamefont {{Pang}},
  \citenamefont {{Tews}}, \citenamefont {{Coughlin}}, \citenamefont {{Bulla}},
  \citenamefont {{Van Den Broeck}},\ and\ \citenamefont {{Dietrich}}}]{Pang21}%
  \BibitemOpen
  \bibfield  {author} {\bibinfo {author} {\bibfnamefont {P.~T.~H.}\
  \bibnamefont {{Pang}}}, \bibinfo {author} {\bibfnamefont {I.}~\bibnamefont
  {{Tews}}}, \bibinfo {author} {\bibfnamefont {M.~W.}\ \bibnamefont
  {{Coughlin}}}, \bibinfo {author} {\bibfnamefont {M.}~\bibnamefont {{Bulla}}},
  \bibinfo {author} {\bibfnamefont {C.}~\bibnamefont {{Van Den Broeck}}}, \
  and\ \bibinfo {author} {\bibfnamefont {T.}~\bibnamefont {{Dietrich}}},\
  }\href {\doibase 10.3847/1538-4357/ac19ab} {\bibfield  {journal} {\bibinfo
  {journal} {Astrophys. J.}\ }\textbf {\bibinfo {volume} {922}},\ \bibinfo
  {eid} {14} (\bibinfo {year} {2021})}\BibitemShut {NoStop}%
\bibitem [{\citenamefont {{Dittmann}}\ \emph {et~al.}(2024)\citenamefont
  {{Dittmann}} \emph {et~al.}}]{Dittmann24}%
  \BibitemOpen
  \bibfield  {author} {\bibinfo {author} {\bibfnamefont {A.~J.}\ \bibnamefont
  {{Dittmann}}} \emph {et~al.},\ }\href {\doibase 10.3847/1538-4357/ad5f1e}
  {\bibfield  {journal} {\bibinfo  {journal} {\apj}\ }\textbf {\bibinfo
  {volume} {974}},\ \bibinfo {eid} {295} (\bibinfo {year} {2024})}\BibitemShut
  {NoStop}%
\bibitem [{\citenamefont {{Rutherford}}\ \emph {et~al.}(2024)\citenamefont
  {{Rutherford}} \emph {et~al.}}]{Rutherford24}%
  \BibitemOpen
  \bibfield  {author} {\bibinfo {author} {\bibfnamefont {N.}~\bibnamefont
  {{Rutherford}}} \emph {et~al.},\ }\href {\doibase 10.3847/2041-8213/ad5f02}
  {\bibfield  {journal} {\bibinfo  {journal} {\apjl}\ }\textbf {\bibinfo
  {volume} {971}},\ \bibinfo {eid} {L19} (\bibinfo {year} {2024})}\BibitemShut
  {NoStop}%
\bibitem [{\citenamefont {Mauviard}\ \emph {et~al.}(2025)\citenamefont
  {Mauviard} \emph {et~al.}}]{Mauviard25}%
  \BibitemOpen
  \bibfield  {author} {\bibinfo {author} {\bibfnamefont {L.}~\bibnamefont
  {Mauviard}} \emph {et~al.},\ }\href {\doibase 10.3847/1538-4357/ae145d}
  {\bibfield  {journal} {\bibinfo  {journal} {\apj}\ }\textbf {\bibinfo
  {volume} {995}},\ \bibinfo {pages} {60} (\bibinfo {year} {2025})}\BibitemShut
  {NoStop}%
\bibitem [{\citenamefont {Bertone}\ and\ \citenamefont
  {Hooper}(2018)}]{Bertone18}%
  \BibitemOpen
  \bibfield  {author} {\bibinfo {author} {\bibfnamefont {G.}~\bibnamefont
  {Bertone}}\ and\ \bibinfo {author} {\bibfnamefont {D.}~\bibnamefont
  {Hooper}},\ }\href {\doibase 10.1103/RevModPhys.90.045002} {\bibfield
  {journal} {\bibinfo  {journal} {Rev. Mod. Phys.}\ }\textbf {\bibinfo {volume}
  {90}},\ \bibinfo {pages} {045002} (\bibinfo {year} {2018})}\BibitemShut
  {NoStop}%
\bibitem [{\citenamefont {Arvanitaki}\ \emph {et~al.}(2010)\citenamefont
  {Arvanitaki}, \citenamefont {Dimopoulos}, \citenamefont {Dubovsky},
  \citenamefont {Kaloper},\ and\ \citenamefont {March-Russell}}]{Arvanitaki10}%
  \BibitemOpen
  \bibfield  {author} {\bibinfo {author} {\bibfnamefont {A.}~\bibnamefont
  {Arvanitaki}}, \bibinfo {author} {\bibfnamefont {S.}~\bibnamefont
  {Dimopoulos}}, \bibinfo {author} {\bibfnamefont {S.}~\bibnamefont
  {Dubovsky}}, \bibinfo {author} {\bibfnamefont {N.}~\bibnamefont {Kaloper}}, \
  and\ \bibinfo {author} {\bibfnamefont {J.}~\bibnamefont {March-Russell}},\
  }\href {\doibase 10.1103/PhysRevD.81.123530} {\bibfield  {journal} {\bibinfo
  {journal} {Phys. Rev. D}\ }\textbf {\bibinfo {volume} {81}},\ \bibinfo
  {pages} {123530} (\bibinfo {year} {2010})}\BibitemShut {NoStop}%
\bibitem [{\citenamefont {Lee}\ \emph {et~al.}(2021)\citenamefont {Lee},
  \citenamefont {Chu},\ and\ \citenamefont {Lin}}]{Lee21}%
  \BibitemOpen
  \bibfield  {author} {\bibinfo {author} {\bibfnamefont {B.~K.~K.}\
  \bibnamefont {Lee}}, \bibinfo {author} {\bibfnamefont {M.-c.}\ \bibnamefont
  {Chu}}, \ and\ \bibinfo {author} {\bibfnamefont {L.-M.}\ \bibnamefont
  {Lin}},\ }\href {\doibase 10.3847/1538-4357/ac2735} {\bibfield  {journal}
  {\bibinfo  {journal} {The Astrophysical Journal}\ }\textbf {\bibinfo {volume}
  {922}},\ \bibinfo {pages} {242} (\bibinfo {year} {2021})}\BibitemShut
  {NoStop}%
\bibitem [{\citenamefont {Colpi}\ \emph {et~al.}(1986)\citenamefont {Colpi},
  \citenamefont {Shapiro},\ and\ \citenamefont {Wasserman}}]{Colpi86}%
  \BibitemOpen
  \bibfield  {author} {\bibinfo {author} {\bibfnamefont {M.}~\bibnamefont
  {Colpi}}, \bibinfo {author} {\bibfnamefont {S.~L.}\ \bibnamefont {Shapiro}},
  \ and\ \bibinfo {author} {\bibfnamefont {I.}~\bibnamefont {Wasserman}},\
  }\href {\doibase 10.1103/PhysRevLett.57.2485} {\bibfield  {journal} {\bibinfo
   {journal} {Phys. Rev. Lett.}\ }\textbf {\bibinfo {volume} {57}},\ \bibinfo
  {pages} {2485} (\bibinfo {year} {1986})}\BibitemShut {NoStop}%
\bibitem [{\citenamefont {{Chavanis}}\ and\ \citenamefont
  {{Harko}}(2012)}]{Chavanis12}%
  \BibitemOpen
  \bibfield  {author} {\bibinfo {author} {\bibfnamefont {P.-H.}\ \bibnamefont
  {{Chavanis}}}\ and\ \bibinfo {author} {\bibfnamefont {T.}~\bibnamefont
  {{Harko}}},\ }\href {\doibase 10.1103/PhysRevD.86.064011} {\bibfield
  {journal} {\bibinfo  {journal} {Phys. Rev. D}\ }\textbf {\bibinfo {volume}
  {86}},\ \bibinfo {eid} {064011} (\bibinfo {year} {2012})}\BibitemShut
  {NoStop}%
\bibitem [{\citenamefont {{Rafiei Karkevandi}}\ \emph
  {et~al.}(2022)\citenamefont {{Rafiei Karkevandi}}, \citenamefont {{Shakeri}},
  \citenamefont {{Sagun}},\ and\ \citenamefont {{Ivanytskyi}}}]{Karkevandi22}%
  \BibitemOpen
  \bibfield  {author} {\bibinfo {author} {\bibfnamefont {D.}~\bibnamefont
  {{Rafiei Karkevandi}}}, \bibinfo {author} {\bibfnamefont {S.}~\bibnamefont
  {{Shakeri}}}, \bibinfo {author} {\bibfnamefont {V.}~\bibnamefont {{Sagun}}},
  \ and\ \bibinfo {author} {\bibfnamefont {O.}~\bibnamefont {{Ivanytskyi}}},\
  }\href {\doibase 10.1103/PhysRevD.105.023001} {\bibfield  {journal} {\bibinfo
   {journal} {Phys. Rev. D}\ }\textbf {\bibinfo {volume} {105}},\ \bibinfo
  {eid} {023001} (\bibinfo {year} {2022})}\BibitemShut {NoStop}%
\bibitem [{\citenamefont {{Eby}}\ \emph {et~al.}(2016)\citenamefont {{Eby}},
  \citenamefont {{Kouvaris}}, \citenamefont {{Nielsen}},\ and\ \citenamefont
  {{Wijewardhana}}}]{Eby16}%
  \BibitemOpen
  \bibfield  {author} {\bibinfo {author} {\bibfnamefont {J.}~\bibnamefont
  {{Eby}}}, \bibinfo {author} {\bibfnamefont {C.}~\bibnamefont {{Kouvaris}}},
  \bibinfo {author} {\bibfnamefont {N.~G.}\ \bibnamefont {{Nielsen}}}, \ and\
  \bibinfo {author} {\bibfnamefont {L.~C.~R.}\ \bibnamefont {{Wijewardhana}}},\
  }\href {\doibase 10.1007/JHEP02(2016)028} {\bibfield  {journal} {\bibinfo
  {journal} {J. High Energy Phys.}\ }\textbf {\bibinfo {volume} {2016}},\
  \bibinfo {eid} {28} (\bibinfo {year} {2016})}\BibitemShut {NoStop}%
\bibitem [{\citenamefont {Li}\ \emph {et~al.}(2012)\citenamefont {Li},
  \citenamefont {Wang},\ and\ \citenamefont {Cheng}}]{XYLi12}%
  \BibitemOpen
  \bibfield  {author} {\bibinfo {author} {\bibfnamefont {X.}~\bibnamefont
  {Li}}, \bibinfo {author} {\bibfnamefont {F.}~\bibnamefont {Wang}}, \ and\
  \bibinfo {author} {\bibfnamefont {K.}~\bibnamefont {Cheng}},\ }\href
  {\doibase 10.1088/1475-7516/2012/10/031} {\bibfield  {journal} {\bibinfo
  {journal} {J. Cosmol. Astropart. Phys.}\ }\textbf {\bibinfo {volume}
  {2012}},\ \bibinfo {pages} {031} (\bibinfo {year} {2012})}\BibitemShut
  {NoStop}%
\bibitem [{\citenamefont {Kouvaris}(2013)}]{kouvaris13}%
  \BibitemOpen
  \bibfield  {author} {\bibinfo {author} {\bibfnamefont {C.}~\bibnamefont
  {Kouvaris}},\ }\href {\doibase https://doi.org/10.1155/2013/856196}
  {\bibfield  {journal} {\bibinfo  {journal} {Adv. High Energy Phys.}\ }\textbf
  {\bibinfo {volume} {2013}},\ \bibinfo {pages} {856196} (\bibinfo {year}
  {2013})}\BibitemShut {NoStop}%
\bibitem [{\citenamefont {Chavanis}(2011)}]{Chavanis11}%
  \BibitemOpen
  \bibfield  {author} {\bibinfo {author} {\bibfnamefont {P.-H.}\ \bibnamefont
  {Chavanis}},\ }\href {\doibase 10.1103/PhysRevD.84.043531} {\bibfield
  {journal} {\bibinfo  {journal} {Phys. Rev. D}\ }\textbf {\bibinfo {volume}
  {84}},\ \bibinfo {pages} {043531} (\bibinfo {year} {2011})}\BibitemShut
  {NoStop}%
\bibitem [{\citenamefont {{Kodama}}\ and\ \citenamefont
  {{Yamada}}(1972)}]{Kodama72}%
  \BibitemOpen
  \bibfield  {author} {\bibinfo {author} {\bibfnamefont {T.}~\bibnamefont
  {{Kodama}}}\ and\ \bibinfo {author} {\bibfnamefont {M.}~\bibnamefont
  {{Yamada}}},\ }\href {\doibase 10.1143/PTP.47.444} {\bibfield  {journal}
  {\bibinfo  {journal} {Prog. Theor. Phys.}\ }\textbf {\bibinfo {volume}
  {47}},\ \bibinfo {pages} {444} (\bibinfo {year} {1972})}\BibitemShut
  {NoStop}%
\bibitem [{\citenamefont {{Comer}}\ \emph {et~al.}(1999)\citenamefont
  {{Comer}}, \citenamefont {{Langlois}},\ and\ \citenamefont
  {{Lin}}}]{Comer99}%
  \BibitemOpen
  \bibfield  {author} {\bibinfo {author} {\bibfnamefont {G.~L.}\ \bibnamefont
  {{Comer}}}, \bibinfo {author} {\bibfnamefont {D.}~\bibnamefont {{Langlois}}},
  \ and\ \bibinfo {author} {\bibfnamefont {L.~M.}\ \bibnamefont {{Lin}}},\
  }\href {\doibase 10.1103/PhysRevD.60.104025} {\bibfield  {journal} {\bibinfo
  {journal} {Phys. Rev. D}\ }\textbf {\bibinfo {volume} {60}},\ \bibinfo {eid}
  {104025} (\bibinfo {year} {1999})}\BibitemShut {NoStop}%
\bibitem [{\citenamefont {{Sandin}}\ and\ \citenamefont
  {{Ciarcelluti}}(2009)}]{Sandin09}%
  \BibitemOpen
  \bibfield  {author} {\bibinfo {author} {\bibfnamefont {F.}~\bibnamefont
  {{Sandin}}}\ and\ \bibinfo {author} {\bibfnamefont {P.}~\bibnamefont
  {{Ciarcelluti}}},\ }\href {\doibase 10.1016/j.astropartphys.2009.09.005}
  {\bibfield  {journal} {\bibinfo  {journal} {Astropart. Phys.}\ }\textbf
  {\bibinfo {volume} {32}},\ \bibinfo {pages} {278} (\bibinfo {year}
  {2009})}\BibitemShut {NoStop}%
\bibitem [{\citenamefont {{Leung}}\ \emph {et~al.}(2012)\citenamefont
  {{Leung}}, \citenamefont {{Chu}},\ and\ \citenamefont {{Lin}}}]{Leung12}%
  \BibitemOpen
  \bibfield  {author} {\bibinfo {author} {\bibfnamefont {S.~C.}\ \bibnamefont
  {{Leung}}}, \bibinfo {author} {\bibfnamefont {M.~C.}\ \bibnamefont {{Chu}}},
  \ and\ \bibinfo {author} {\bibfnamefont {L.~M.}\ \bibnamefont {{Lin}}},\
  }\href {\doibase 10.1103/PhysRevD.85.103528} {\bibfield  {journal} {\bibinfo
  {journal} {Phys. Rev. D}\ }\textbf {\bibinfo {volume} {85}},\ \bibinfo {eid}
  {103528} (\bibinfo {year} {2012})}\BibitemShut {NoStop}%
\bibitem [{\citenamefont {{Kain}}(2020)}]{Kain20}%
  \BibitemOpen
  \bibfield  {author} {\bibinfo {author} {\bibfnamefont {B.}~\bibnamefont
  {{Kain}}},\ }\href {\doibase 10.1103/PhysRevD.102.023001} {\bibfield
  {journal} {\bibinfo  {journal} {\prd}\ }\textbf {\bibinfo {volume} {102}},\
  \bibinfo {eid} {023001} (\bibinfo {year} {2020})}\BibitemShut {NoStop}%
\bibitem [{\citenamefont {{Henriques}}\ \emph {et~al.}(1990)\citenamefont
  {{Henriques}}, \citenamefont {{Liddle}},\ and\ \citenamefont
  {{Moorhouse}}}]{Henriques90b}%
  \BibitemOpen
  \bibfield  {author} {\bibinfo {author} {\bibfnamefont {A.~B.}\ \bibnamefont
  {{Henriques}}}, \bibinfo {author} {\bibfnamefont {A.~R.}\ \bibnamefont
  {{Liddle}}}, \ and\ \bibinfo {author} {\bibfnamefont {R.~G.}\ \bibnamefont
  {{Moorhouse}}},\ }\href {\doibase 10.1016/0370-2693(90)90789-9} {\bibfield
  {journal} {\bibinfo  {journal} {Phys. Lett. B}\ }\textbf {\bibinfo {volume}
  {251}},\ \bibinfo {pages} {511} (\bibinfo {year} {1990})}\BibitemShut
  {NoStop}%
\bibitem [{\citenamefont {{Hippert}}\ \emph {et~al.}(2023)\citenamefont
  {{Hippert}}, \citenamefont {{Dillingham}}, \citenamefont {{Tan}},
  \citenamefont {{Curtin}}, \citenamefont {{Noronha-Hostler}},\ and\
  \citenamefont {{Yunes}}}]{Hippert23}%
  \BibitemOpen
  \bibfield  {author} {\bibinfo {author} {\bibfnamefont {M.}~\bibnamefont
  {{Hippert}}}, \bibinfo {author} {\bibfnamefont {E.}~\bibnamefont
  {{Dillingham}}}, \bibinfo {author} {\bibfnamefont {H.}~\bibnamefont {{Tan}}},
  \bibinfo {author} {\bibfnamefont {D.}~\bibnamefont {{Curtin}}}, \bibinfo
  {author} {\bibfnamefont {J.}~\bibnamefont {{Noronha-Hostler}}}, \ and\
  \bibinfo {author} {\bibfnamefont {N.}~\bibnamefont {{Yunes}}},\ }\href
  {\doibase 10.1103/PhysRevD.107.115028} {\bibfield  {journal} {\bibinfo
  {journal} {\prd}\ }\textbf {\bibinfo {volume} {107}},\ \bibinfo {eid}
  {115028} (\bibinfo {year} {2023})},\ \Eprint
  {http://arxiv.org/abs/2211.08590} {arXiv:2211.08590 [astro-ph.HE]}
  \BibitemShut {NoStop}%
\bibitem [{\citenamefont {Sun}\ and\ \citenamefont {Wen}(2024)}]{Sun24}%
  \BibitemOpen
  \bibfield  {author} {\bibinfo {author} {\bibfnamefont {H.}~\bibnamefont
  {Sun}}\ and\ \bibinfo {author} {\bibfnamefont {D.}~\bibnamefont {Wen}},\
  }\href {\doibase 10.1103/PhysRevD.109.123037} {\bibfield  {journal} {\bibinfo
   {journal} {Phys. Rev. D}\ }\textbf {\bibinfo {volume} {109}},\ \bibinfo
  {pages} {123037} (\bibinfo {year} {2024})}\BibitemShut {NoStop}%
\bibitem [{\citenamefont {Ellis}\ \emph {et~al.}(2018)\citenamefont {Ellis}
  \emph {et~al.}}]{Ellis18}%
  \BibitemOpen
  \bibfield  {author} {\bibinfo {author} {\bibfnamefont {J.}~\bibnamefont
  {Ellis}} \emph {et~al.},\ }\href {\doibase 10.1103/PhysRevD.97.123007}
  {\bibfield  {journal} {\bibinfo  {journal} {Phys. Rev. D}\ }\textbf {\bibinfo
  {volume} {97}},\ \bibinfo {pages} {123007} (\bibinfo {year}
  {2018})}\BibitemShut {NoStop}%
\bibitem [{\citenamefont {Ivanytskyi}\ \emph {et~al.}(2020)\citenamefont
  {Ivanytskyi}, \citenamefont {Sagun},\ and\ \citenamefont
  {Lopes}}]{Ivanytskyi20}%
  \BibitemOpen
  \bibfield  {author} {\bibinfo {author} {\bibfnamefont {O.}~\bibnamefont
  {Ivanytskyi}}, \bibinfo {author} {\bibfnamefont {V.}~\bibnamefont {Sagun}}, \
  and\ \bibinfo {author} {\bibfnamefont {I.}~\bibnamefont {Lopes}},\ }\href
  {\doibase 10.1103/PhysRevD.102.063028} {\bibfield  {journal} {\bibinfo
  {journal} {Phys. Rev. D}\ }\textbf {\bibinfo {volume} {102}},\ \bibinfo
  {pages} {063028} (\bibinfo {year} {2020})}\BibitemShut {NoStop}%
\bibitem [{\citenamefont {Hartle}(1967)}]{Hartle67}%
  \BibitemOpen
  \bibfield  {author} {\bibinfo {author} {\bibfnamefont {J.~B.}\ \bibnamefont
  {Hartle}},\ }\href {\doibase 10.1086/149400} {\bibfield  {journal} {\bibinfo
  {journal} {Astrophys. J.}\ }\textbf {\bibinfo {volume} {150}},\ \bibinfo
  {pages} {1005} (\bibinfo {year} {1967})}\BibitemShut {NoStop}%
%%CITATION = ASJOA,150,1005;%%
\bibitem [{\citenamefont {Binnington}\ and\ \citenamefont
  {Poisson}(2009)}]{Binnington09}%
  \BibitemOpen
  \bibfield  {author} {\bibinfo {author} {\bibfnamefont {T.}~\bibnamefont
  {Binnington}}\ and\ \bibinfo {author} {\bibfnamefont {E.}~\bibnamefont
  {Poisson}},\ }\href {\doibase 10.1103/PhysRevD.80.084018} {\bibfield
  {journal} {\bibinfo  {journal} {Phys. Rev. D}\ }\textbf {\bibinfo {volume}
  {80}},\ \bibinfo {pages} {084018} (\bibinfo {year} {2009})}\BibitemShut
  {NoStop}%
%%CITATION = ARXIV:0906.1366;%%
\bibitem [{\citenamefont {Damour}\ and\ \citenamefont
  {Nagar}(2010)}]{Damour10}%
  \BibitemOpen
  \bibfield  {author} {\bibinfo {author} {\bibfnamefont {T.}~\bibnamefont
  {Damour}}\ and\ \bibinfo {author} {\bibfnamefont {A.}~\bibnamefont {Nagar}},\
  }\href {\doibase 10.1103/PhysRevD.81.084016} {\bibfield  {journal} {\bibinfo
  {journal} {Phys. Rev. D}\ }\textbf {\bibinfo {volume} {81}},\ \bibinfo
  {pages} {084016} (\bibinfo {year} {2010})}\BibitemShut {NoStop}%
%%CITATION = ARXIV:0911.5041;%%
\bibitem [{\citenamefont {Leung}\ \emph {et~al.}(2022)\citenamefont {Leung},
  \citenamefont {Chu},\ and\ \citenamefont {Lin}}]{Leung22}%
  \BibitemOpen
  \bibfield  {author} {\bibinfo {author} {\bibfnamefont {K.-L.}\ \bibnamefont
  {Leung}}, \bibinfo {author} {\bibfnamefont {M.-C.}\ \bibnamefont {Chu}}, \
  and\ \bibinfo {author} {\bibfnamefont {L.-M.}\ \bibnamefont {Lin}},\ }\href
  {\doibase 10.1103/PhysRevD.105.123010} {\bibfield  {journal} {\bibinfo
  {journal} {Phys. Rev. D}\ }\textbf {\bibinfo {volume} {105}},\ \bibinfo
  {pages} {123010} (\bibinfo {year} {2022})}\BibitemShut {NoStop}%
\bibitem [{\citenamefont {{Zhang}}\ \emph {et~al.}(2023)\citenamefont
  {{Zhang}}, \citenamefont {{Luo}}, \citenamefont {{Tsao}}, \citenamefont
  {{Chen}},\ and\ \citenamefont {{Lin}}}]{Zhang23}%
  \BibitemOpen
  \bibfield  {author} {\bibinfo {author} {\bibfnamefont {K.}~\bibnamefont
  {{Zhang}}}, \bibinfo {author} {\bibfnamefont {L.-W.}\ \bibnamefont {{Luo}}},
  \bibinfo {author} {\bibfnamefont {J.-S.}\ \bibnamefont {{Tsao}}}, \bibinfo
  {author} {\bibfnamefont {C.-S.}\ \bibnamefont {{Chen}}}, \ and\ \bibinfo
  {author} {\bibfnamefont {F.-L.}\ \bibnamefont {{Lin}}},\ }\href {\doibase
  10.1016/j.rinp.2023.106967} {\bibfield  {journal} {\bibinfo  {journal}
  {Results Phys.}\ }\textbf {\bibinfo {volume} {53}},\ \bibinfo {eid} {106967}
  (\bibinfo {year} {2023})}\BibitemShut {NoStop}%
\bibitem [{\citenamefont {Flanagan}\ and\ \citenamefont
  {Hinderer}(2008)}]{Flanagan08}%
  \BibitemOpen
  \bibfield  {author} {\bibinfo {author} {\bibfnamefont {E.~E.}\ \bibnamefont
  {Flanagan}}\ and\ \bibinfo {author} {\bibfnamefont {T.}~\bibnamefont
  {Hinderer}},\ }\href {\doibase 10.1103/PhysRevD.77.021502} {\bibfield
  {journal} {\bibinfo  {journal} {Phys. Rev. D}\ }\textbf {\bibinfo {volume}
  {77}},\ \bibinfo {pages} {021502(R)} (\bibinfo {year} {2008})}\BibitemShut
  {NoStop}%
\bibitem [{\citenamefont {Hinderer}(2008)}]{Hinderer08}%
  \BibitemOpen
  \bibfield  {author} {\bibinfo {author} {\bibfnamefont {T.}~\bibnamefont
  {Hinderer}},\ }\href {\doibase 10.1086/533487} {\bibfield  {journal}
  {\bibinfo  {journal} {Astrophys. J.}\ }\textbf {\bibinfo {volume} {677}},\
  \bibinfo {pages} {1216} (\bibinfo {year} {2008})}\BibitemShut {NoStop}%
%%CITATION = ARXIV:0711.2420;%%
\bibitem [{\citenamefont {Lyne}\ \emph {et~al.}(2004)\citenamefont {Lyne} \emph
  {et~al.}}]{Lyne04}%
  \BibitemOpen
  \bibfield  {author} {\bibinfo {author} {\bibfnamefont {A.~G.}\ \bibnamefont
  {Lyne}} \emph {et~al.},\ }\href {\doibase 10.1126/science.1094645} {\bibfield
   {journal} {\bibinfo  {journal} {Science}\ }\textbf {\bibinfo {volume}
  {303}},\ \bibinfo {pages} {1153} (\bibinfo {year} {2004})}\BibitemShut
  {NoStop}%
\bibitem [{\citenamefont {Lattimer}\ and\ \citenamefont
  {Schutz}(2005)}]{Lattimer05}%
  \BibitemOpen
  \bibfield  {author} {\bibinfo {author} {\bibfnamefont {J.~M.}\ \bibnamefont
  {Lattimer}}\ and\ \bibinfo {author} {\bibfnamefont {B.~F.}\ \bibnamefont
  {Schutz}},\ }\href {\doibase 10.1086/431543} {\bibfield  {journal} {\bibinfo
  {journal} {Astrophys. J.}\ }\textbf {\bibinfo {volume} {629}},\ \bibinfo
  {pages} {979} (\bibinfo {year} {2005})}\BibitemShut {NoStop}%
\bibitem [{\citenamefont {Kramer}\ and\ \citenamefont {Wex}(2009)}]{Kramer09}%
  \BibitemOpen
  \bibfield  {author} {\bibinfo {author} {\bibfnamefont {M.}~\bibnamefont
  {Kramer}}\ and\ \bibinfo {author} {\bibfnamefont {N.}~\bibnamefont {Wex}},\
  }\href {\doibase 10.1088/0264-9381/26/7/073001} {\bibfield  {journal}
  {\bibinfo  {journal} {Classical and Quantum Gravity}\ }\textbf {\bibinfo
  {volume} {26}},\ \bibinfo {pages} {073001} (\bibinfo {year}
  {2009})}\BibitemShut {NoStop}%
\bibitem [{\citenamefont {{Hartle}}\ and\ \citenamefont
  {{Thorne}}(1968)}]{Hartle68}%
  \BibitemOpen
  \bibfield  {author} {\bibinfo {author} {\bibfnamefont {J.~B.}\ \bibnamefont
  {{Hartle}}}\ and\ \bibinfo {author} {\bibfnamefont {K.~S.}\ \bibnamefont
  {{Thorne}}},\ }\href {\doibase 10.1086/149707} {\bibfield  {journal}
  {\bibinfo  {journal} {\apj}\ }\textbf {\bibinfo {volume} {153}},\ \bibinfo
  {pages} {807} (\bibinfo {year} {1968})}\BibitemShut {NoStop}%
\bibitem [{\citenamefont {{Goldman}}\ \emph {et~al.}(2013)\citenamefont
  {{Goldman}}, \citenamefont {{Mohapatra}}, \citenamefont {{Nussinov}},
  \citenamefont {{Rosenbaum}},\ and\ \citenamefont {{Teplitz}}}]{Goldman13}%
  \BibitemOpen
  \bibfield  {author} {\bibinfo {author} {\bibfnamefont {I.}~\bibnamefont
  {{Goldman}}}, \bibinfo {author} {\bibfnamefont {R.~N.}\ \bibnamefont
  {{Mohapatra}}}, \bibinfo {author} {\bibfnamefont {S.}~\bibnamefont
  {{Nussinov}}}, \bibinfo {author} {\bibfnamefont {D.}~\bibnamefont
  {{Rosenbaum}}}, \ and\ \bibinfo {author} {\bibfnamefont {V.}~\bibnamefont
  {{Teplitz}}},\ }\href {\doibase 10.1016/j.physletb.2013.07.017} {\bibfield
  {journal} {\bibinfo  {journal} {Phys. Lett. B}\ }\textbf {\bibinfo {volume}
  {725}},\ \bibinfo {pages} {200} (\bibinfo {year} {2013})}\BibitemShut
  {NoStop}%
\bibitem [{\citenamefont {{Kouvaris}}\ and\ \citenamefont
  {{Nielsen}}(2015)}]{Kouvaris15}%
  \BibitemOpen
  \bibfield  {author} {\bibinfo {author} {\bibfnamefont {C.}~\bibnamefont
  {{Kouvaris}}}\ and\ \bibinfo {author} {\bibfnamefont {N.~G.}\ \bibnamefont
  {{Nielsen}}},\ }\href {\doibase 10.1103/PhysRevD.92.063526} {\bibfield
  {journal} {\bibinfo  {journal} {Phys. Rev. D}\ }\textbf {\bibinfo {volume}
  {92}},\ \bibinfo {eid} {063526} (\bibinfo {year} {2015})}\BibitemShut
  {NoStop}%
\bibitem [{\citenamefont {{Di Giovanni}}\ \emph {et~al.}(2020)\citenamefont
  {{Di Giovanni}}, \citenamefont {{Fakhry}}, \citenamefont {{Sanchis-Gual}},
  \citenamefont {{Degollado}},\ and\ \citenamefont {{Font}}}]{DiGiovanni20}%
  \BibitemOpen
  \bibfield  {author} {\bibinfo {author} {\bibfnamefont {F.}~\bibnamefont {{Di
  Giovanni}}}, \bibinfo {author} {\bibfnamefont {S.}~\bibnamefont {{Fakhry}}},
  \bibinfo {author} {\bibfnamefont {N.}~\bibnamefont {{Sanchis-Gual}}},
  \bibinfo {author} {\bibfnamefont {J.~C.}\ \bibnamefont {{Degollado}}}, \ and\
  \bibinfo {author} {\bibfnamefont {J.~A.}\ \bibnamefont {{Font}}},\ }\href
  {\doibase 10.1103/PhysRevD.102.084063} {\bibfield  {journal} {\bibinfo
  {journal} {Phys. Rev. D}\ }\textbf {\bibinfo {volume} {102}},\ \bibinfo {eid}
  {084063} (\bibinfo {year} {2020})}\BibitemShut {NoStop}%
\end{thebibliography}%

\end{document}